\documentclass[pre,aps]{revtex4}
\usepackage{graphicx}
\usepackage{amsfonts}
\usepackage{amssymb}
\usepackage{amsmath}
\usepackage{psfrag}
\usepackage{natbib}
\usepackage{mathtools}
\usepackage{subfigure,float}
\usepackage{epsfig}
\usepackage{epstopdf}
\usepackage[colorlinks=true, linkcolor=blue, urlcolor=blue, citecolor=black]{hyperref}
\newcommand{\beqa}{\begin{eqnarray}}
\newcommand{\eeqa}{\end{eqnarray}}
\newcommand{\beq}{\begin{equation}}
\newcommand{\eeq}{\end{equation}}
\newcommand{\grad}{{\boldsymbol \nabla}}

\begin{document}
\title{
  The relationship between viscoelasticity and elasticity
     }

\author{
J. H. Snoeijer$^1$ and A. Pandey$^1$ and M. A. Herrada$^2$ and J. Eggers$^3$
}
\affiliation{
$^1$Physics of Fluids Group, Faculty of Science and Technology,
Mesa+ Institute, University of Twente, 7500 AE Enschede, The Netherlands \\
$^2$ Depto. de Mec\'anica de Fluidos e Ingenier\'ia Aeroespacial,
Universidad de Sevilla, E-41092 Sevilla, Spain. \\
$^3$School of Mathematics,
University of Bristol, University Walk,
Bristol BS8 1TW, United Kingdom
}

\begin{abstract}
We consider models for elastic liquids, such as solutions of flexible
polymers. They introduce a relaxation time $\lambda$ into the system,
over which stresses relax. We study the kinematics of the problem,
and clarify the relationship between Lagrangian and Eulerian descriptions,
thereby showing which polymer models correspond to a nonlinear elastic
deformation in the limit $\lambda\rightarrow\infty$. This allows us to
split the change in elastic energy into reversible and dissipative parts,
and thus to write an equation for the total energy, the sum of kinetic
and elastic energies. As an illustration, we show how the presence
or absence of an elastic limit determines the fate of an elastic thread
during capillary instability, using novel numerical schemes based on
our insights into the flow kinematics. 
\end{abstract}
\maketitle

\section{Introduction}

Complex fluids, possessing a characteristic time scale $\lambda$ much
larger than the relaxation time of a simple fluid are extremely common
and important \cite{BAH87,Larson99,MS15}. On one hand, they serve to
explain and interpret the behavior of a vast range of biological and
industrial fluids. On the other hand they are also a key to understanding
a range of new and fascinating instabilities \cite{CEFLM06,MvS07}.

The key to describing viscoelastic fluids is the separation of time
scales between for example the relaxation time of a polymer, which can
be seconds \cite{CEFLM06}, and microscopic relaxation times, which
are many orders of magnitude smaller. As a result, there is an
expectation that there is an order parameter (or slow variable)
which is able to describe the extra degrees of freedom associated
with the microstructure \cite{MPP72,Leonov1976,  BE_book,Oe_book}.
The resulting
continuum descriptions have to be kinematically consistent, in that
they should not change their form upon a change of coordinate systems.
In addition, the description must be consistent with requirements of
local thermodynamic equilibrium, and elaborate formalisms have been
developed to ensure this \cite{BE_book,Oe_book}.

In formulating equations of motion for a viscoelastic fluid, one can either
begin with the equations for a simple fluid, and introduce memory into
the dynamics of a fluid element. Alternatively, one can start from
the equations for an elastic system, and introduce a fading memory of
the original reference state \cite{BookTanner}.
This alternative is rarely followed
within the fluids community, but has the advantage that it explicitly
relates to elasticity theory. In this paper we expand on this alternative,
using a formulation that resembles that by Leonov~\cite{Leonov1976},
and clear up some of the confusion which surrounded
the elastic limit in the past \cite{TPLB00,BGKO01,MLPB16b}.
For a range of popular viscoelastic models we will analyze the
elastic limit, thereby clarifying issues which have remained obscure
in the past. For example, the so-called the upper and lower convected
derivatives represent different ways of computing the time derivative of
a tensor embedded in a velocity field in a consistent fashion \cite{BAH87};
we explain how the two choices describe physically different situations.

With the elastic limit and the corresponding expression for the stress in
hand, we are able to separate two fundamentally different aspects of
viscoelasticity: the relationship between the stress and the state of the
complex fluid (polymer) on one hand, and the dynamics of relaxation from
the deformed state, on the other. The former governs the reversible change
in elastic energy upon deformation, the latter the dissipative part.
In particular, we are able to construct an energy balance equation in
a systematic fashion, where the total energy is made up of the kinetic and
elastic energies. The energy dissipation is determined by the relaxation
dynamics of the order parameter, which is known as the conformation tensor.

\begin{figure}
\centering
\includegraphics[width=0.7\hsize]{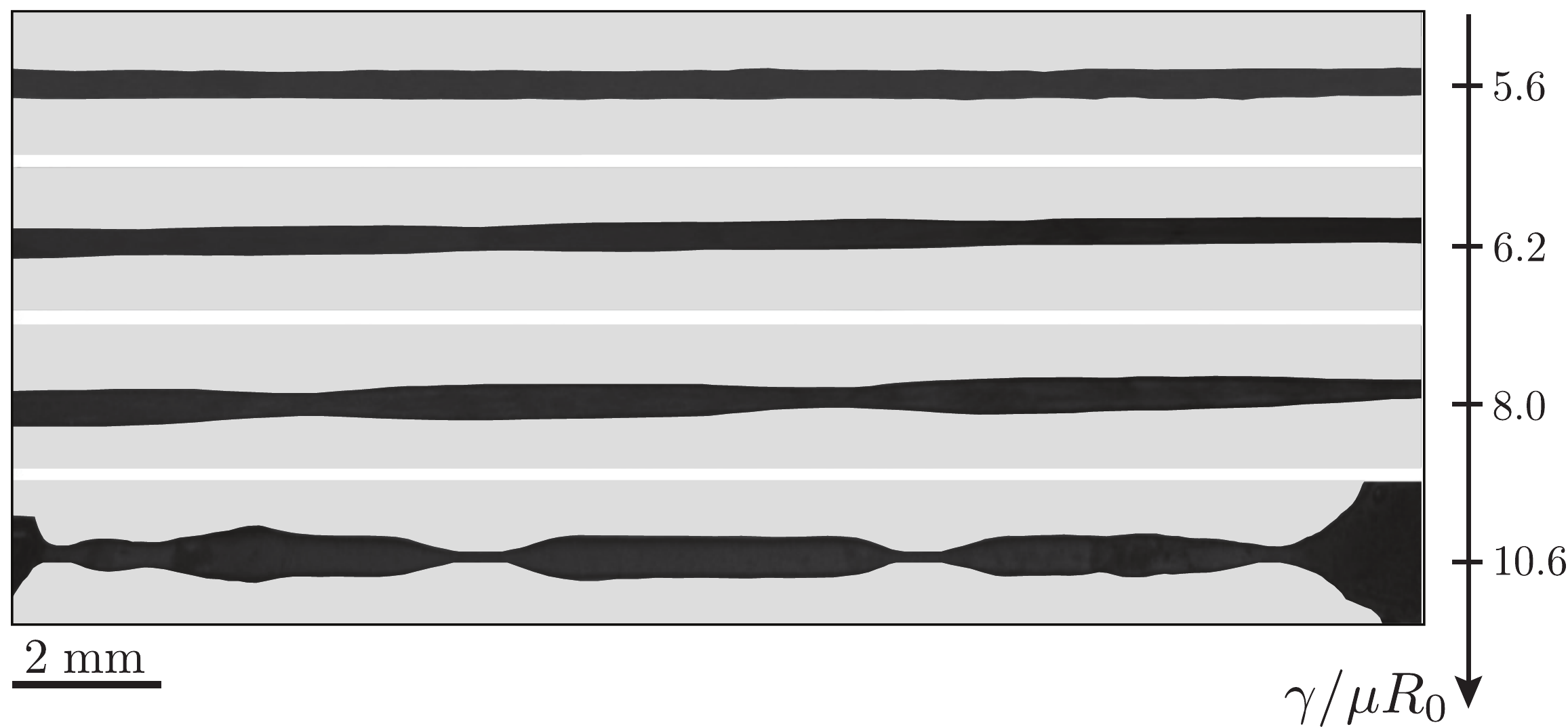}
\caption{ Equilibrium shape of agar gel cylinders for different
values of the shear modulus. The radius is 240 $\mu$m, the surface
tension is $\gamma$ = 37 mN/m. Adapted from \cite{MPFPP10}.
}
\label{fig:intro}
\end{figure}

Our findings are important to recent developments in the mechanics of
exceedingly soft elastic
solids~\cite{AndreottiSoftMatt2016,StyleDufresneReview,MPFPP10}. As an
example we quote the capillary instability of an elastic
cylinder~\cite{EYa84,MPFPP10,EF_book,XB17}, as shown in Fig.~\ref{fig:intro}.
The cylinder consists of a fully cross-linked agar gel, and unlike viscoelastic
liquids therefore exhibits a reference state that is free from elastic
stress -- one could call it an elastic solid rather than an elastic
liquid. When sufficiently soft, however, the solid cylinder still exhibits
a Rayleigh-Plateau instability that leads to the formation of thin threads.
This resembles a bit the ``bead-on-a-string" configuration, which is a
well-known feature for thinning jets of dilute polymer suspensions
\cite{CEFLM06}. A natural question is whether the two systems can be
described by the same continuum description.

The paper starts with a brief overview of the continuum formulations of
elastic liquids and elastic solids in Section~\ref{sec:classical}.
Particular emphasis
is given to the kinematics of affine and non-affine motions in the limit
$\lambda \rightarrow \infty$. Based on this, we propose in
Section~\ref{sec:order} an order parameter formulation for elastic liquids,
which is similar in spirit to the work by Leonov~\cite{Leonov1976}.
We derive the energy equation and provide a systematic classification
of polymer models -- the most common models are summarized in
Appendix~\ref{app:rheo}.  At the end of the paper we present a numerical
study of the capillary instability of elastic threads, highlighting
the importance of the elastic limit (Section~\ref{sec:collapse}) and
we close with a discussion (Section~\ref{sec:discussion}).

\section{Classical continuum theory}
\label{sec:classical}
\subsection{Viscoelastic fluids}

We consider an incompressible velocity field ${\bf v}({\bf x},t)$
($\grad\cdot{\bf v}=0$), described by the momentum balance
\beq
\rho\left(\frac{\partial{\bf v}}{\partial t} +
{\bf v}\cdot\grad{\bf v}\right) =
\grad\cdot\boldsymbol{\sigma},
\label{NS}
\eeq
where $\boldsymbol{\sigma}$ is the stress tensor.
The stress tensor is split into a Newtonian contribution (coming e.g. from
a solvent) and a polymeric (viscoelastic) contribution $\boldsymbol{\sigma}_p$:
\beq
\boldsymbol{\sigma} = -p{\bf I} +
\eta_s \dot{\boldsymbol{\gamma}} +
\boldsymbol{\sigma}_p \equiv -p{\bf I} + \boldsymbol{\tau},
\label{sigma_OB}
\eeq
where the deviatoric stress $\boldsymbol{\tau}$
is the contribution excluding the
pressure. Here we defined the rate-of-deformation tensor
\beq
\boldsymbol{\dot{\gamma}} = \left(\grad{\bf v}\right) +
\left(\grad{\bf v}\right)^T,
\eeq
and $\eta_s$ is the solvent viscosity. Any isotropic contribution to
the stress can be written as part of the pressure $p$.

The non-Newtonian contribution $\boldsymbol{\sigma}_p$ originates from
the stretching of the microstructure inside the fluid. Though we will refer to
$\boldsymbol{\sigma}_p$ as the ``polymeric stress",
having in mind dilute polymer
suspensions, the concept equally applies to emulsions whose microstructure
is described by droplet deformations \cite{MWB17,MWB18}. 
The non-Newtonian contribution $\boldsymbol{\sigma}_p$ is governed by
a separate evolution equation, describing the state of the component
governed by a long time scale $\lambda$, for example a polymer.
We will see it is best to write the polymeric stress
$\boldsymbol{\sigma}_p = \boldsymbol{\sigma}_p({\bf A})$ as a
function of a state variable (or order parameter field) ${\bf A}$ that
is a symmetric second rank tensor. This conformation tensor, which can be
derived by coarse-graining microscopic models based on suspended
beads-and-spring dumbbells \cite{BAH87}, has the property that the polymer
stress vanishes when $\mathbf A=\mathbf I$, where $\mathbf I$ is the
identity tensor. However, in this paper we follow a purely continuum approach,
without referring to any microscopic model. In particular, we will give a
precise continuum definition of the conformation tensor, and show that
the structure of the dependence $\boldsymbol{\sigma}_p({\bf A})$ will
be determined by the elastic limit $\lambda\rightarrow\infty$.

In the simplest case this relationship is linear,
\beq
\boldsymbol{\sigma}_p = \mu\left(\mathbf A - \mathbf I \right),
\label{neo_Hooke}
\eeq
where $\mu$ has the dimensions of an elastic shear modulus.
The state variable ${\bf A}$ must have the property that it evolves
toward its relaxed state $\mathbf A=\mathbf I$ in the limit of long times.
Once more, the simplest way of doing this to write a linear relaxation law
\[
\dot{{\bf A}} = -\frac{1}{\lambda}\left({\bf A}-{\bf I}\right).
\]
However, it is has been known for a long time \cite{O50,BAH87} that
for the dynamics of a second-rank tensor to be {\it frame invariant}
\cite{MS15}, i.e. to be independent of the frame of reference, the ordinary
time derivative needs to be replaced by another derivative. There are
two versions of this frame invariant derivative (and linear combinations
thereof). The so-called {\it upper convected derivative}
\beq
\overset{\triangledown}{\bf A} =
\frac{\partial{\bf A}}{\partial t} +
{\bf v}\cdot\grad{\bf A}
- \left(\grad{\bf v}\right)^T \cdot {\bf A} -
{\bf A}\cdot\left(\grad{\bf v}\right),
\label{ucd}
\eeq
is derived from the requirement that it transforms consistently as a
covariant tensor. The first two terms on the right are the convected
derivative of a material point, ensuring Galilean invariance, the
last two terms make sure that ${\bf A}$ transforms correctly under
deformations by the flow.
We will see below that the upper convected derivative describes
{\it affine} motion, that is a situation where each material point
of the polymer follows the flow. This is seen most directly from
beads-and-spring models of polymers \cite{BAH87}, from which
the upper convected derivative follows automatically if beads are required
to follow the flow without slip. However, a contravariant
formulation would do equally well from the point of view of frame invariance,
but yields a different derivative, known as the lower convected derivative:
\beq
\overset{\vartriangle}{\bf A} =
\frac{\partial{\bf A}}{\partial t} + {\bf v}\cdot\grad{\bf A}
+ {\bf A}\cdot\left(\grad{\bf v}\right)^T + \left(\grad{\bf v}\right)\cdot{\bf A}.
\label{lcd}
\eeq

Using the upper convected derivative,
a linear relaxation law takes the form
\begin{equation}
\label{eq:alambda}
\overset{\triangledown}{\mathbf A} =
- \frac{1}{\lambda} \left( \mathbf A - \mathbf I \right),
\end{equation}
which means that for any initial condition $\mathbf A$ relaxes
exponentially toward $\mathbf A = \mathbf I$ on a time scale
$\lambda$. Equations \eqref{neo_Hooke} and \eqref{eq:alambda} can
be combined into a single equation of motion for the polymeric stress:
\beq
\boldsymbol{\sigma}_p +
\lambda \overset{\triangledown}{\boldsymbol{\sigma}}_p
= \eta_p \boldsymbol{\dot{\gamma}},
\label{ucmm}
\eeq
which is a tensorial form of the simple Maxwell fluid; here
$\eta_p=\mu\lambda$ is the polymeric viscosity. The stress
tensor \eqref{sigma_OB}
together with \eqref{ucmm} is known as the Oldroyd B model \cite{BAH87};
in the limit of vanishing rates of deformation, it describes a Newtonian
fluid of total viscosity $\eta_0 = \eta_s + \eta_p$.

Although the Oldroyd B model is very popular owing to its simplicity,
there are many relevant physical effects which are not captured. For that
reason there are many extensions of the Oldroyd B equations, for example
taking into account non-linearity in both \eqref{neo_Hooke} and
\eqref{eq:alambda}, or in the solvent contribution in \eqref{sigma_OB}.
In Appendix~\ref{app:rheo}, we supply a list of various
models. Apart from the question of frame invariance, models have
to be consistent with requirements of thermodynamics
\cite{BM94,MWB17,MWB18,BE_book,Oe_book}.

\subsection{Elasticity}

While fluid mechanics is usually formulated in an Eulerian frame,
non-linear (finite deformation) elasticity is written in a Lagrangian
formulation. This means that deformations are described by a mapping
${\bf x} = {\bf x}({\bf X})$, where ${\bf x}$ is the position of a material
point after the deformation, which used to be at ${\bf X}$ before the
deformation. Elasticity is based on the idea that stresses in an
isotropic medium can only depend on the change in distance between material
points generated by a deformation \cite{LL7,Wiki_notes_FST}.
Namely if
${\bf F}=\partial \mathbf x/\partial \mathbf X$ is the deformation gradient
tensor, and $ds$ is the distance between two points which used to be
a distance $dS$ apart, we obtain \cite{LL7,Wiki_notes_FST}, using
$d{\bf x} = {\bf F}\cdot d{\bf X}$, that
\beq
ds^2 - dS^2 = d\mathbf X^T \cdot
\left( \mathbf F^T \cdot \mathbf F - {\bf I}\right) \cdot d\mathbf X.
\label{dsdS}
\eeq
Thus the energy of a deformation can only depend on Green's deformation
tensor ${\bf C} = \mathbf F^T \cdot \mathbf F$, which is a symmetric second
rank tensor that is defined on the reference configuration
(the tensor $({\bf F}^T\cdot {\bf F} - {\bf I})/2$ is called the finite
strain tensor). The eigenvalues of $\mathbf C$ represent the principal
stretches, i.e. the ratio of deformed over reference length along the
principle directions of the deformation. Importantly, $\mathbf C$ shares
the same eigenvalues as those of the Finger tensor \cite{Wiki_notes_FST},
defined as $\mathbf B = \mathbf F \cdot \mathbf F^T$. The Finger tensor
is entirely defined on the current configuration, and is therefore more
convenient when connecting to the Eulerian descriptions, as typically
used for viscoelastic liquids.
Since the elastic energy can only depend on invariants of ${\bf C}$, which
are the same as those of ${\bf B}$, we can write $W = W({\bf B})$
for the elastic free energy density. Moreover, the function $W({\bf B})$
must have the property that it assumes a minimum for ${\bf B}={\bf I}$.
This means that any deformation will cost energy, which is a necessary
condition for the unstressed state to be stable.

Though not necessary, we from now on focus on incompressible media for which
$\det\left({\mathbf F}\right)=1$. Once the free energy is specified,
the (true) stress tensor for incompressible media follows as
\begin{equation}
\label{eq:virtualwork}
\boldsymbol{\sigma}_p =
\frac{\partial W}{\partial \mathbf F} \cdot \mathbf F^T
=  2 \frac{\partial W}{\partial \mathbf B} \cdot \mathbf B,
\end{equation}
where in the second step we exploited the symmetry of $\mathbf B$.
This expression is a consequence of the virtual work principle
\cite{TN_book}, requiring that any change of the elastic energy
density satisfies
\beq
\frac{dW}{dt} = \boldsymbol{\sigma}_p : (\nabla \mathbf v)^T =
\frac{1}{2} \boldsymbol{\sigma}_p : \dot{\boldsymbol{\gamma}}
\label{reversible}
\eeq
\cite{LL7}, where in the second step we used the symmetry of
$\boldsymbol{\sigma}_p$. The derivation of
\eqref{eq:virtualwork},\eqref{reversible} will be spelled out in
the next section. As we argued before, $W$ can only be a function
of one of
the invariants of ${\bf B}$, which can be written as \cite{Wiki_notes_FST}
\beq
I_1 = B_{kk}, \quad \quad I_2 = \frac{1}{2}\left( B_{kk}^2 - B_{ij}B_{ij} \right),
\quad \quad I_3 = \det \mathbf B,
\label{B_inv}
\eeq
where we remind that $I_3 = \det({\bf F})^2 = 1$ for incompressible media.
Hence,  we can write the free energy as a function of the first two invariants only: $W(I_1,I_2)$.
The constraint $I_3=1$ will be ensured by an isotropic pressure acting as a Lagrange multiplier.

Using \eqref{eq:virtualwork}, we obtain
\begin{equation}
\boldsymbol{\sigma}_p =  2 W_1 \mathbf B +
2W_2\left( {\rm tr}(\mathbf B) \mathbf B - \mathbf B \cdot \mathbf B\right),
\end{equation}
where $W_1\equiv\partial W/\partial I_1$ and
$W_2\equiv\partial W/\partial I_2$. This can be simplified using the
Cayley-Hamilton theorem, which for $\det({\bf B})=1$ reads
\beq
{\bf B}^{-1} = {\bf B}^{2} - {\rm tr}({\bf B}){\bf B} + \frac{1}{2}
\left({\rm tr}({\bf B})^2 - {\rm tr}({\bf B}^2)\right) \mathbf I.
\label{CH}
\eeq
As a result, the stress can be written as
\begin{equation}
\label{eq:sigmaelastic}
\boldsymbol{\sigma}_p =  2 W_1 \left( \mathbf B - \mathbf I \right) +
2 W_2 \left( \mathbf I - \mathbf B^{-1}  \right),
\end{equation}
where for convenience we have absorbed an isotropic contribution
into the pressure.

The derivatives $W_1$ and $W_2$ can be arbitrary non-linear
functions of the invariants $I_1$ and $I_2$. If $W_1=\mu/2$ and
$W_2=0$, one finds the neo-Hookean model, and using \eqref{eq:sigmaelastic}
one finds the stress to be
$\boldsymbol{\sigma}_p = \mu\left(\mathbf B - \mathbf I \right)$,
which corresponds to \eqref{neo_Hooke} with ${\bf A}\equiv{\bf B}$.
The case where both $W_1$ and $W_2$ are constant but non-zero goes
by the name of Mooney-Rivlin solid.

\subsection{Kinematics: affine and non-affine motion}
\label{sub:kinematics}
To make a connection between the Eulerian polymer models and
the Lagrangian formulation of elasticity, we need to find out what
are the deformations generated by transport by the velocity field
${\bf v}$. The velocity field is connected to the motion of material
points by  ${\bf v} = d\mathbf x/dt$, where $d/dt$ is a time derivative
at constant material point $\mathbf X$. Then it follows from the chain
rule that \cite{MS15}
\beq
\frac{d{\bf F}}{dt} = \left(\grad{\bf v}\right)^T\cdot {\bf F}, \quad
\frac{d{\bf F^{-1}}}{dt} = -{\bf F}^{-1}\cdot \left(\grad{\bf v}\right)^T,
\label{F_der}
\eeq
where $\left(\grad{\bf v}\right)_{ij} = \partial_i v_j$. The relation
\eqref{F_der} permits to calculate the deformation gradient tensor from
${\bf v}$, and thus to pass from an Eulerian to a Lagrangian description.
For later reference, we will derive a number of auxiliary kinematic relations. 
The first is obtained by taking the time derivative of the identity
${\bf x}\left({\bf X}({\bf x},t),t\right) = {\bf x}$ at constant ${\bf x}$.
Denoting the derivative at constant $\mathbf x$ by $\partial/\partial t$
and at constant ${\bf X}$ as $d/dt$, we have
$0 = d{\bf x}/dt + \partial \mathbf x/\partial
\mathbf X\cdot \partial{\bf X}/\partial t$,
and so the velocity ${\bf v} = d{\bf x}/dt$ can be calculated from
\beq
{\bf v} = -{\bf F}\cdot \frac{\partial{\bf X}}{\partial t}.
\label{velocity}
\eeq
Thus given the inverse Lagrangian map (or reference map \cite{Kamrin2012})
${\bf X}({\bf x},t)$, the Eulerian velocity field can be retrieved.
Note that \eqref{velocity} can also be written in the form 
\beq
\frac{\partial {\bf X}}{\partial t} + {\bf v}\cdot\grad{\bf X} = 0,
\label{reference}
\eeq
which expresses the fact that by definition the total derivative
of the reference state ${\bf X}$ must vanish. 

If the fluid is incompressible, the mapping
${\bf x}\left({\bf X},t\right)$ must be volume-preserving and so
the Eulerian incompressibility condition $\grad\cdot{\bf v}=0$ is
equivalent to $\det{\bf F}=1$. In the more general case of a compressible
fluid, using the transformation formula for volumes, the Lagrangian
condition simply changes to 
\beq
\rho = \frac{\rho_0}{\det({\bf F})}, 
\label{copressible_L}
\eeq
where $\rho_0$ is the density in reference coordinates. To confirm
that in an Eulerian transformation, 
we take the convected time derivative of \eqref{copressible_L}:
\[
\frac{\partial \rho}{\partial t} + ({\bf v}\cdot\grad)\rho = \frac{d{\rho}}{dt} = 
\rho_0\frac{d{\det({\bf F}^{-1})}}{dt} =
\rho_0{\rm tr}\left(\frac{\bf F}{\det({\bf F})}\frac{d{\bf F}^{-1}}{dt}\right)=
-\frac{\rho_0}{\det({\bf F})}{\rm tr}\left(\grad{\bf v}\right) =
-\rho \grad\cdot{\bf v},
\]
having used Jacobi's formula \cite{Magnus_Neudecker} in the third step,
and \eqref{F_der} in the fourth. In other words, \eqref{copressible_L}
is equivalent to
\beq
\frac{\partial \rho}{\partial t} + \grad\cdot(\rho{\bf v}) = 0,
\label{copressible_E}
\eeq
the usual Eulerian form of the continuity equation. 

Now we turn to the main point of interest, dealing with frame invariant
time derivatives of an Eulerian tensor ${\bf A}(\mathbf x,t)$, and
relating it to the Lagrangian mapping. To achieve that, we derive
the following identity relating the upper convective derivative and
the time derivatives of the mapping:
\beq
\overset{\triangledown}{\bf A} = {\bf F}\cdot
\left[\frac{d}{dt}\left({\bf F}^{-1}\cdot{\bf A}
  \cdot{\bf F}^{-T}\right)\right]\cdot{\bf F}^T.
\label{up_relation}
\eeq
Indeed, making use of \eqref{F_der}, the explicit evaluation of the time
derivative gives (\ref{ucd}).

The definition (\ref{up_relation}) has a natural interpretation. Since
convection plays no role on the domain of material coordinates, one first
projects the Eulerian tensor $\mathbf A$ back to the Lagrangian domain,
using the inverse transformation ${\bf F}^{-1}\cdot{\bf A}\cdot{\bf F}^{-T}$. Then,
once the time-derivative is performed on the Lagrangian domain, the
result is returned to the Eulerian domain to yield an objective tensorial
time-derivative. However, the above procedure is not unique. Namely,
we can also construct a Lagrangian tensor as ${\bf F}^{T}\cdot{\bf A}\cdot{\bf F}$.
This can be viewed as the transformation of a covariant tensor
(while ${\bf F}^{-1}\cdot{\bf A}\cdot{\bf F}^{-T}$ gives the transformation of
a contravariant tensor) \cite{MS15}.
Following the same procedure as above, this gives an alternative time
derivative
\beq
\overset{\vartriangle}{\bf A} = {\bf F}^{-T}\cdot
\left[\frac{d}{dt}\left({\bf F}^T\cdot{\bf A}\cdot{\bf F}\right)\right]\cdot{\bf F}^{-1},
\label{down_relation}
\eeq
where the lower convective derivative is defined in \eqref{lcd}.

\subsubsection{Affine motion}

This produces the sought-after connection between polymer dynamics
and elasticity: in the limit $\lambda\rightarrow\infty$ the derivative
vanishes, so the upper convected models reduce to $\overset{\triangledown}{\bf A} = 0$. Thus multiplying
\eqref{up_relation} by ${\bf F}^{-1}$ from the left and ${\bf F}^{-T}$
from the right, and integrating over time, we find
\beq
{\bf F}^{-1}\cdot{\bf A}\cdot{\bf F}^{-T} = {\rm constant} \quad
\Rightarrow \quad {\bf A}= {\bf F} \cdot \mathbf{D}_0 \cdot {\bf F}^{T},
\eeq
where $\mathbf D_0$ is a constant (time-independent) Lagrangian tensor
that is independent of the mapping; $\mathbf D_0$ can be viewed as an
integration constant and can be determined from initial conditions.
A natural choice is to consider the initial condition ($\mathbf F=\mathbf I$) to be stress-free,
i.e. $\boldsymbol{\sigma}_p=0$ which implies $\mathbf A=\mathbf I$.
Hence, we find $\mathbf D_0=\mathbf I$, i.e.
$\mathbf A = {\bf B} \equiv \mathbf F \cdot \mathbf F^T$.

In other words, given a relaxation law of the form
$\lambda\overset{\triangledown}{\bf A} = f({\bf A})$, in the limit
$\lambda\rightarrow\infty$ the upper convected derivative implies that
${\bf A}$ tracks the stretching induced by the flow, as
described by \eqref{dsdS}: the deformation follows the flow affinely.
The conformation tensor $\mathbf A$ in viscoelasticity plays the same
role as the Finger tensor $\mathbf B$ in elasticity.
If on the other hand the relaxation law would imply
$\overset{\vartriangle}{\bf A} = 0$, using \eqref{down_relation} we
find $\mathbf A = {\bf B}^{-1} \equiv \mathbf F^{-T} \cdot\mathbf F^{-1}$,
which would correspond to an elastic response in a direction opposite the flow.
Let us illustrate these two cases using the simple elongational flow
\beq
v_r = -\dot{\epsilon} r, \quad v_z = 2\dot{\epsilon} z.
\label{elong}
\eeq
Integrating \eqref{F_der} with initial condition ${\bf F}={\bf I}$
one obtains
\beq
\mathbf B =
\begin{pmatrix}
e^{-2\dot{\epsilon}t} & 0 \\
0 & e^{4\dot{\epsilon}t} &
\end{pmatrix}, \quad
{\mathbf B}^{-1} =
\begin{pmatrix}
e^{2\dot{\epsilon}t} & 0 \\
0 & e^{-4\dot{\epsilon}t} &
\end{pmatrix}.
\label{B_elong}
\eeq
In other words, ${\bf B}$ describes stretching in the $z$-direction
and contraction in the radial direction that is generated by the flow $\mathbf v$, while ${\bf B}^{-1}$ does
the exact opposite.

\subsubsection{Non-affine motion}

By combining the upper and lower derivatives, one can describe a
situation where the polymer deformation partially follows the flow,
making it non-affine to a certain degree. To show this, we consider the
derivative introduced in the Johnson-Segalman model \cite{JS77}
which takes into account the possibility that the polymer does not
follow the flow of the solvent in an affine fashion, but slips
with respect to the flow. This is accomplished by
introducing the polymer velocity $\mathbf v_a$, which satisfies
\begin{equation}\label{eq:slip}
\grad{\bf v}_a = \frac{a}{2} \left[  \grad{\bf v} + (\grad{\bf v})^T \right]
 + \frac{1}{2} \left[  \grad{\bf v} - (\grad{\bf v})^T \right]
= \frac{1+a}{2}   \grad{\bf v} - \frac{1-a}{2} (\grad{\bf v})^T,
\end{equation}
where $a$ (the so-called slip parameter) satisfies $-1\le a\le 1$.
Indeed, for $a=1$, $\grad{\bf v}_a$ and $\grad{\bf v}$ are the same,
and the polymer follows perfectly. This is no longer the case for $a\neq 1$.
The antisymmetric part of $\grad{\bf v}_a$ and $\grad{\bf v}$ are the same, which means ${\bf v}_a$ and ${\bf v}$ have
the same vorticity, so that the polymer follows any solid body rotation
of the flow perfectly. On the other hand, the rate of deformation of
the polymer (symmetric part) satisfies
$\dot{\boldsymbol{\gamma}}_a = a \dot{\boldsymbol{\gamma}}$.
The Johnson-Segalman derivative is the upper convected derivative
with respect to the slipping polymer:
\begin{eqnarray}\label{eq:slipderivative}
\overset{\triangledown}{\left(\bf A\right)}_a  &\equiv&
\frac{d{\mathbf{A}}}{d t}
- \left(\grad{\bf v}_a\right)^T\cdot{\mathbf{A}} -
{\mathbf{A}}\cdot\left(\grad{\bf v}_a\right)
= \frac{1+a}{2}\overset{\triangledown}{\mathbf A}
+ \frac{1-a}{2}\overset{\vartriangle}{\mathbf A}.
\end{eqnarray}
Here we remind that in the following $d/dt$ denotes the material derivative.

To illustrate the consequences of the non-affine motion, we consider
a relaxation law based on the Johnson-Segalman derivative. In that
case, one obtains
$\overset{\triangledown}{\left(\bf A\right)}_a  = 0 $ in the limit of
large relaxation times. We solve this equation for a uniform
shear flow $\mathbf v = \dot{\gamma} y \mathbf e_x$, and take the
initial conditions as $\mathbf A=\mathbf I$.
Solving for ${\bf A}$, one obtains the oscillatory response:
\beq
{\bf A} =
\begin{pmatrix}
\frac{1}{(1-a)}\left[1 - a
  \cos\left(\sqrt{1-a^2} \dot{\gamma} t\right)\right] &
\frac{a}{\sqrt{1-a^2}}
\sin\left(\sqrt{1-a^2} \dot{\gamma} t\right) \\
\frac{a}{\sqrt{1-a^2}}
\sin\left(\sqrt{1-a^2} \dot{\gamma} t\right) &
\frac{1}{(1+a)}\left[1 + a
\cos\left(\sqrt{1-a^2} \dot{\gamma} t\right)\right]
\end{pmatrix}.
\eeq
Clearly, this does not correspond to any elastic model.
Physically, the oscillations
can be understood from the non-affine kinematics dictated by (\ref{eq:slip}).
The flow $\mathbf v = \dot{\gamma} y \mathbf e_x$ can be written as a
superposition of an elongational flow and a rigid body rotation, which
are exactly equal of amplitude. Any slip ($a<1$) removes part of the
elongation flow, while the full rigid body rotation is retained. This
effectively leads to an ``excess" rigid body motion, that gives rise to
a periodic ``flow" of the polymer, with a frequency
$\sqrt{1-a^2}\dot{\gamma}$. We remark that these
oscillations have a purely kinematic origin, and thus persist for a
Johnson-Segalman fluid that is sheared at finite values of $\lambda$.
In that case, however, the oscillations are damped so that $\mathbf A$,
and hence the polymer stress, eventually reaches a steady state \cite{Larson99}.

From these observations we draw an important conclusion.
When the order parameter $\mathbf A$ tracks the stretching of the polymer,
just like the Finger tensor $\mathbf B$ does in the theory of elasticity,
the relaxation law must be constructed from the upper convected derivative.
Only then, one can make sure that the polymer deformation is purely
elastic in the limit $\lambda \rightarrow \infty$, in the sense that
it follows the deformation imposed by the flow. Any other time-derivative
implies non-affine motion. The very same conclusions were reached in the
context of emulsions, whose drops deform into ellipsoids -- in that case
the eigenvalues of $\mathbf A$ represent the square of the semi-axes of
the deformed droplets \cite{MWB17,MWB18}. 

\section{Order parameter formulation of rheological models}
\label{sec:order}
Using arguments similar to those of \cite{Leonov1976}, now we want to
combine the above observations, in order to formulate
a class of models which separate the description of viscoelastic stress
into two different aspects; as shown in more detail in
Appendix~\ref{app:curvi}, the eigenvalues of the conformation tensor
$\mathbf A$ represent the stretching of the polymer. The
first aspect concerns the energetics of the problem, from which one can
calculate the stress.
The second aspect describes the relaxation of ${\bf A}$, from which
we calculate the dissipation. Taken together, this provides us with a
systematic procedure to find the equation describing the total energy.
To summarize, we have the following ingredients:
\begin{enumerate}
\item A symmetric rank-2 tensor order parameter field
$\mathbf A(\mathbf x,t)$, which quantifies the stretched state of the polymer.
\item An elastic free energy density $W(\mathbf A)$, which is minimal
  for $\mathbf A = \mathbf I$.
\item A relaxation equation towards $\mathbf A=\mathbf I$,
  governing dissipation.
\end{enumerate}
The physical structure we are trying to embody in $\mathbf A$, using
these requirements, has been characterized as an
``instantaneous reference state" in \cite{BookTanner}. At each instant,
if the flow were to stop, the instantaneous reference state tends
to the current state. In terms of the curvilinear formalism
(cf. Appendix~\ref{app:curvi}), the conformation tensor relates to the
difference between the ``instantaneous reference metric", and the actual
``current metric". The former relaxes to the latter.

To find the correct structure of $W(\mathbf A)$, we borrow from
the elastic energy $W(\mathbf B)$, as found on non-linear elasticity.
The energy must be a function of the invariants
\beq
I_1 = A_{kk}, \quad \quad I_2 = \frac{1}{2}\left( A_{kk}^2 - A_{ij}A_{ij} \right),
\quad \quad I_3 = \det \mathbf A,
\label{A_inv}
\eeq
with ${\bf A}$ taking the role of ${\bf B}$ in \eqref{B_inv}. There
should be no confusion from using the same notation for the invariants
of ${\bf A}$.
The choice of $W(\mathbf A)$ naturally determines the elastic limit,
while the relaxation equation for $\mathbf A$ accounts for irreversible
dissipation.

We now proceed to derive the expression for the stress and the dissipation,
focusing first on \emph{affine} polymer models, for which the conformation
tensor relaxes according to
$\lambda\overset{\triangledown}{\bf A} = f({\bf A})$. The idea is that the
reversible part of the deformation has the same form as the reversible change in free energy
\eqref{reversible}, so the remainder corresponds to dissipation.
Writing the work in symmetric form
$\frac{1}{2} \boldsymbol{\sigma}_p : \dot{\boldsymbol{\gamma}}$, and
introducing the volumetric dissipation rate $\epsilon_p$, we find
\beq
\label{eq:energdissip}
\frac{1}{2} \boldsymbol{\sigma}_p : \dot{\boldsymbol{\gamma}}  =
\frac{dW}{dt} + \epsilon_p.
\end{equation}
This expresses that any work done during the deformation must either be
stored in elastic energy, or be dissipated. With this convention,
$\epsilon_p$ must be positive in order to be consistent with
thermodynamics. The time derivative $dW/dt$ can be calculated using the
definition of $\overset{\triangledown}{\bf A}$, yielding
\begin{eqnarray}
  \frac{dW}{dt}  &=& \frac{\partial W}{\partial \mathbf A}:
  \frac{d \mathbf A}{dt}
= \frac{\partial W}{\partial \mathbf A}: \left[
  \left(\grad{\bf v}\right)^T \cdot \mathbf A + \mathbf A \cdot
  \left(\grad{\bf v}\right) + \overset{\triangledown} {\bf A} \right]
 \nonumber \\
 &=&
 \left(\frac{\partial W}{\partial \mathbf A} \cdot \mathbf A \right)
 : \dot{\boldsymbol{\gamma}}
 + \frac{\partial W}{\partial \mathbf A}:
 \overset{\triangledown}{\bf A},
\label{eq:hups}
\end{eqnarray}
where in the last line we made use of the symmetry $\mathbf A$.
As anticipated in (\ref{eq:energdissip}), this nicely separates into a
term due to deformation $\dot{\boldsymbol{\gamma}}$ and to a term
associated to the relaxation law. Equating $dW/dt$ in \eqref{eq:hups}
and \eqref{eq:energdissip}, we obtain the expression for stress.
\beq
\boldsymbol{\sigma}_p = 2\frac{\partial W}{\partial \mathbf A} \cdot \mathbf A.
\label{sigma_A}
\eeq
As expected, this is exactly the form of the elastic stress
\eqref{eq:virtualwork},
with ${\bf A}$ replacing ${\bf B}$. The remainder can be identified as
the dissipation
\begin{equation}
\epsilon_p = -\frac{\partial W}{\partial \mathbf A}:
\overset{\triangledown}{\bf A}.
\label{P}
\end{equation}

We are now in a position to formulate the energy balance for a polymeric
liquid. Multiplying \eqref{NS} by ${\bf v}$, using \eqref{sigma_OB},
we obtain
\beq
\frac{1}{2}\frac{\partial \rho v^2}{\partial t} +
\grad\cdot\left[\left(\frac{\rho v^2}{2}+p \right){\bf v}
-\eta_s\dot{\boldsymbol{\gamma}} \cdot {\bf v}  -
\boldsymbol{\sigma}_p \cdot {\bf v}\right]
= - \epsilon -\frac{1}{2}\boldsymbol{\sigma}_p:\dot{\boldsymbol{\gamma}},
\label{NS_energy}
\eeq
where
\beq
\epsilon = \frac{\eta_s}{2}\dot{\boldsymbol{\gamma}}:\dot{\boldsymbol{\gamma}}
\label{epsilon}
\eeq
is the viscous dissipation. Using
$\boldsymbol{\sigma}_p:\dot{\boldsymbol{\gamma}}/2=dW/dt + \epsilon_p$
this can be rewritten as
\beq
\frac{d}{d t}\left(\frac{\rho v^2}{2} + W\right) +
\grad\cdot\left[p {\bf v}
-\eta_s\dot{\boldsymbol{\gamma}} \cdot {\bf v}  -
\boldsymbol{\sigma}_p \cdot {\bf v}\right] = - \epsilon - \epsilon_p,
\label{NS_energy_elas}
\eeq
which has the form of a conservation law for the sum of kinetic
energy $\rho v^2/2$ and elastic energy $W$. The term in square brackets is the
energy flux. The right hand side represents the dissipation, which has a
viscous contribution from the solvent $\epsilon$, and a polymeric
contribution $\epsilon_p$, which
according to \eqref{P} is associated with the relaxation of ${\bf A}$.
The conservation law \eqref{NS_energy_elas}, together with the expressions
for the stress \eqref{sigma_A} and the dissipation \eqref{P}, are the main
results of this paper.

In the remainder we restrict ourselves to incompressible order parameters
for which $\det(\mathbf A)=1$ (this restriction could be lifted if necessary),
which makes $W$ a function of $I_1,I_2$ only: $W(I_1,I_2)$.
Evaluating the elastic stress (\ref{sigma_A}) is a essentially a repeat of the
elastic calculation \eqref{eq:virtualwork},\eqref{eq:sigmaelastic},
and $\boldsymbol{\sigma}_p$ has the same form:
\beq
\label{eq:stressfromenergy}
\boldsymbol{\sigma}_p =  2 W_1 \left(\mathbf A-\mathbf I\right) +
2 W_2 \left(\mathbf I - \mathbf A^{-1} \right),
\eeq
but is based on the conformation tensor $\mathbf A$ rather than on the
Finger tensor $\mathbf B$. The relaxation
of the conformation tensor gives rise to dissipation $\epsilon_p$, which using
$W(I_1,I_2)$ and \eqref{P} becomes
\beq
\label{eq:p}
\epsilon_p = -\left[W_1 \mathbf I + W_2 {\rm tr}(\mathbf A) \mathbf I - W_2
    \mathbf A \right] : \overset{\triangledown}{{\mathbf{A}}}.
\eeq
It is evident that dissipation vanishes in the absence of relaxation
$\overset{\triangledown}{{\mathbf{A}}}=0$; in this case
$\mathbf A = \mathbf B$, so that we recover the full structure of
elasticity theory.

As an aside, we note that \eqref{eq:p} can be written in a more elegant
form, using the lower convected derivative. Namely, using the identity
\[
{\rm tr}({\bf A}) {\rm tr}(\overset{\triangledown}{\bf A}) -
{\bf A}:\overset{\triangledown}{\bf A} =
{\rm tr}(\overset{\vartriangle}{\left(\mathbf A^{-1}\right)}),
\]
we can also write $\epsilon_p$ in the form
\beq
\label{eq:p_simp}
\epsilon_p = -W_1 {\rm tr} (\overset{\triangledown}{\bf A}) -
W_2{\rm tr} \overset{\vartriangle}{\left(\mathbf A^{-1} \right)},
\eeq
so that the full energy balance takes the form
\beq
\frac{d}{d t} \left( \frac{\rho v^2}{2} + W \right) -
\grad\cdot\left[ \boldsymbol{\sigma} \cdot {\bf v}\right] = - \epsilon
-W_1 tr (\overset{\triangledown}{\bf A}) -
W_2 tr \overset{\vartriangle}{\left(\mathbf A^{-1} \right)}.
\label{energy_two}
\eeq

\subsection{Examples}
\label{sub:ex}
Let us discuss a number of viscoelastic models that are captured
by the present approach. A more detailed list is found in
Appendix~\ref{app:rheo}. As regards the validity of a certain model,
the subtle question concerns the relaxation equation for ${\bf A}$,
which here we take of the form
$\lambda\overset{\triangledown}{\bf A} = f({\bf A})$, so that
deformations follows the flow affinely. Non-affine motion will be
considered below. In addition, it has to be checked that the model
is consistent with thermodynamics and that in particular,
$\epsilon_p >0$.

First, we consider the Oldroyd B model, where relaxation equation is
known as the upper convected Maxwell model. The model is defined by
a neo-Hookean elastic energy, while the relaxation is linear, i.e.
\beq
W=\frac{\mu}{2}\left(I_1-3\right), \quad
\overset{\triangledown}{{\mathbf{A}}}=
-\frac{1}{\lambda}\left( \mathbf A - \mathbf I \right).
\label{W_Hooke}
\eeq
Hence, the stress (\ref{eq:stressfromenergy}) and dissipation
(\ref{eq:p}) follow as \cite{HS_les_houches}:
\beq
\boldsymbol{\sigma}_p = \mu\left( \mathbf A - \mathbf I \right),
\quad \epsilon_p =  \frac{\mu(I_1 - 3)}{2\lambda} = \frac{W}{\lambda}.
\label{sigma_P_Hooke}
\eeq
Note that since $W$ must be positive (it acquires its minimum for
${\bf A} = {\bf I}$ where it is zero), this implies that
$\epsilon_p \ge 0$, as required.

Second, in both the energy and the relaxation equation,
there can appear nonlinear
functions of the invariants. The most popular of such models is the
FENE-P model \cite{BAH87,Larson99}. Like other models of the same
kind, it is based on the concept of an elastic spring attached to two
beads in solution. While a Hookean, non-interacting spring leads
to the Oldroyd B equation,
here the spring is non-linear, so that it cannot be extended beyond a
limiting length $L$. This avoids the deficiency of the Oldroyd B
model, that the polymeric stress grows exponentially to infinity in a
strong flow. In the non-linear case the model can no longer be solved
exactly, so various approximations are used of which FENE-P is one.
The finite extensibility is introduced so that
$I_1 \equiv {\rm tr}(\mathbf A)$ reaches a maximum value $L^2$, via
the stress relation
\beq
\boldsymbol{\sigma}_p =
\mu F(I_1) \left( \mathbf A - \mathbf I \right), \quad \mathrm{with} \quad
F(I_1)= \frac{L^2-3}{L^2-I_1}.
\eeq
Clearly, the stress diverges when $I_1=L^2$. This stress relation is complemented by a nonlinear relaxation law

\beq
\overset{\triangledown}{\mathbf A} = - \frac{1}{\lambda}
\left( F(I_1)\mathbf A - \mathbf I \right),\label{eq:feneprelax}
\eeq
which reduces to $\overset{\triangledown}{\mathbf A} =0$ for infinite
relaxation time. It is now an easy task to identify the rubber model that
emerges in the elastic limit. Namely, from (\ref{eq:stressfromenergy})
we can find the elastic energy as
\begin{equation}
W  = \frac{\mu}{2}(L^2-3) \ln\left(F(I_1) \right),
\end{equation}
which in rubber elasticity is known as the Gent model. Using that
in the elastic limit $\mathbf A = \mathbf F \cdot \mathbf F^T $ and
$I_1={\rm tr}(\mathbf F \cdot \mathbf F^T)$, the corresponding stress reads
\beq
\boldsymbol{\sigma}_p = \mu F(I_1)
\left( \mathbf F\cdot \mathbf F^T  - \mathbf I \right).
\eeq
This shows how the viscoelastic FENE-P model is naturally connected
to the Gent model. We remark that the FENE-CR model \cite{CR88} has the
same energetic structure as the FENE-P model, but with a slightly
different relaxation law, namely
$\overset{\triangledown}{\mathbf A} = - F(I_1)
\left( \mathbf A - \mathbf I \right)/\lambda$.

More generally, models that only involve the first invariant
$I_1$ must have a relaxation equation of the form
\begin{equation}
\overset{\triangledown}{\mathbf A} = - \frac{1}{\lambda}\left( g_1(I_1) \mathbf A - g_0(I_1) \mathbf I\right),
\end{equation}
since any multiples of the type $\mathbf A \cdot \mathbf A$ would, upon taking the trace, produce a second invariant $I_2$. Hence, the dissipation for $I_1$-based models becomes
\begin{equation}
\epsilon_p =  \frac{W_1}{\lambda}\left[ I_1 g_1(I_1) - 3 g_0(I_1)\right].
\end{equation}
Given that $W_1 >0$, thermodynamic consistency $\epsilon \geq 0$ is then ensured by
\begin{equation}
I_1 g_1(I_1) - 3 g_0(I_1) \geq 0 \quad \mathrm{for} \quad I_1 \geq 3.
\end{equation}
The FENE-P and FENE-CR models indeed fall within this class.

\subsection{Non-affine models}
\label{sub:non-affine}
So far we have dealt with the physical situation that the constituents
follow the flow exactly. As discussed in Subsection~\ref{sub:kinematics},
using a derivative which is a linear superposition of upper and lower
convected derivatives, one can model a situation where the material
``slips'' relative to the flow. In that case, dissipative processes are
described by the relaxation equation in the slipping frame, leading to a
form $\lambda\overset{\triangledown}{\left(\bf A\right)}_a = f({\bf A})$,
where $\overset{\triangledown}{\left(\bf A\right)}_a$ is defined in
\eqref{eq:slipderivative}, which is the upper convected derivative in
the polymer frame.

In order to split $dW/dt$ into a part that depends on the deformation
and a part that depends on the relaxation, we
use $\overset{\triangledown}{\left(\bf A\right)}_a$ instead of
$\overset{\triangledown}{\bf A}$ to obtain
\begin{eqnarray}
\frac{dW}{dt}  = \frac{\partial W}{\partial \mathbf A}: \frac{d \mathbf A}{dt}
= \left(a\frac{\partial W}{\partial \mathbf A} \cdot \mathbf A \right)
: \dot{\boldsymbol{\gamma}}
+ \frac{\partial W}{\partial \mathbf A}:
\overset{\triangledown}{\left(\bf A\right)}_a.
\end{eqnarray}
We therefore find the stress and dissipation, respectively, as
\beq
\boldsymbol{\sigma}_p = 2a \frac{\partial W}{\partial \mathbf A}
\cdot \mathbf A, \quad \quad
\epsilon_p = -\frac{\partial W}{\partial \mathbf A}:
\overset{\triangledown}{\left( \bf A \right)}_a,
\label{non-affine_diss}
\eeq
where $\boldsymbol{\sigma}_p$ has the same form as \eqref{eq:virtualwork},
but with a factor $a$ in front of the expression for the stress. This
reflects the slip: the polymer is stretched less than expected, making
the response ``softer" by a fraction $a$. By consequence, the stress
can be further expressed as
\begin{equation}
\label{eq:stressfromenergy_na}
\boldsymbol{\sigma}_p =  2 a W_1 \left(\mathbf A-\mathbf I\right)
+ 2 a W_2 \left(\mathbf I - \mathbf A^{-1} \right).
\end{equation}

A Johnson-Segalman fluid is characterized by a simple neo-Hookean energy, $W_1=\mu/2$ and
$W_2=0$, complemented by a linear relaxation equation in the slipping frame
\beq
\label{eq:JSscaled}
\overset{\triangledown}{\left( \bf A \right)}_a =
\frac{1+a}{2}\overset{\triangledown}{\bf A} +
\frac{1-a}{2}\overset{\vartriangle}{\bf A}
= - \frac{1}{\lambda}\left( \mathbf A - \mathbf I\right).
\eeq
Hence instead of \eqref{sigma_P_Hooke} we have
\beq
\boldsymbol{\sigma}_p = a\mu\left( \mathbf A - \mathbf I \right),
\quad \epsilon_p =  \frac{\mu(I_1 - 3)}{2\lambda} = \frac{W}{\lambda}.
\label{sigma_P_Hooke2}
\eeq
In particular, it follows that the dissipation is always positive,
and vanishes in the limit $\lambda\rightarrow\infty$. Indeed,
one verifies that (\ref{eq:JSscaled}) with $\lambda =\infty$ is invariant
under the symmetry $t \mapsto -t$, $\mathbf v \mapsto - \mathbf v$. So
upon reversing the flow, the system dynamics retraces its path and there is
no relaxation of $\mathbf A$ in the frame co-moving with the polymer.
Remarkably, however, we have seen that owing to non-affine motion,
there is an oscillatory response to a shear deformation.
By consequence, even though the ``elastic limit" is non-dissipative and
invariant upon time reversal, it does not correspond to any elastic solid.
Perhaps this is related to the fact that within the formalism developed
by Beris and Edwards \cite{BE_book}, the Johnson-Segalman model has a
contribution from the dissipation bracket, even at infinite relaxation time,
indicating aspects of irreversible behavior. 

As a particular example we consider the case of perfect counterflow, $a=-1$,
with a neo-Hookean energy for the polymer, $W=\frac{1}{2}\mu(I_1-3)$.
According to the above, we recover the lower convected Maxwell model
\beq
\boldsymbol{\sigma}_p = -\mu\left( \mathbf A - \mathbf I \right), \quad
\overset{\vartriangle}{\bf A} =
-\frac{1}{\lambda}\left( \mathbf A - \mathbf I \right),
\label{lc_Maxwell}
\eeq
which can be written as
\beq
\boldsymbol{\sigma}_p +
\lambda \overset{\vartriangle}{\boldsymbol{\sigma}}_p =
\eta_p \boldsymbol{\dot{\gamma}},
\label{lcmm}
\eeq
as opposed to \eqref{ucmm} for the upper convected version.
We previously concluded that the in corresponding elastic limit, since
$\overset{\vartriangle}{\bf A} = 0$, we had $\mathbf A = \mathbf B^{-1}$.
This gives the elastic stress
\[
\boldsymbol{\sigma}_p = \mu\left({\bf I} - {\bf B}^{-1}\right).
\]
Comparing to (\ref{eq:sigmaelastic}), we find that this corresponds to a
rubber model with an energy $W = \frac{1}{2}\mu(I_2^{\bf B}-3)$, based on
the \emph{second} invariant of the Finger tensor,  $I^{\mathbf B}_2$.
This is consistent with a neo-Hookean energy for $\mathbf A$, since
using the Cayley-Hamilton theorem (\ref{CH}) one verifies that
$I^{\mathbf B}_2= I^{\mathbf B^{-1}}_1=I^{\mathbf A}_1$ for incompressible media.

In general, we thus conclude that any non-affine motion $a\neq 1$ does
not converge to an elastic-solid response in the limit of large relaxation
time. Formally, on the level of stress, the special case of perfect
counterflow \emph{does} converge to a special type of elastic solid
that is based on the second invariant of $\mathbf B$. However, no
realistic rubber is described only by the second invariant of the
deformation \cite[p. 158]{TN_book},~\cite{Mihai17}. This can be rationalized by the
fact that the polymer needs to deform oppositely to the imposed flow.

\section{Collapse of a cylinder under surface tension}
\label{sec:collapse}
\begin{figure}
\centering
\includegraphics[width=0.7\hsize]{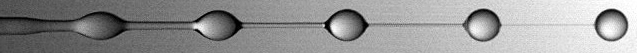}
\caption{High-speed video-image of a jet (radius 0.3 mm)
of dilute (0.01 wt\%) aqueous polyacrylamide solution undergoing
capillary thinning \cite{CEFLM06}. The relaxation time of the polymer
is $\lambda = 0.012$ s.
}
\label{fig:jet}
\end{figure}

We now illustrate our observations by considering a more complex
example including a free surface: the breakup of a liquid column
under the action of surface tension \cite{CEFLM06,EF_book,TLEAD18}.
This is illustrated in Fig.~\ref{fig:jet}, showing the breakup of a
water jet containing a low concentration of a high molecular flexible
polymer, with a relaxation time of about 0.01 seconds. The breakup
process repeats itself periodically, so one can see an almost cylindrical
thread at different stages of thinning. An alternative geometry is that
of a liquid bridge between two plates, which leads to a single thread.
In each case, the thread radius is observed to thin
exponentially \cite{BELR97,CEFLM06,TLEAD18}, with a rate of decay
$1/(3\lambda)$ in the case of a single timescale.

Using a lubrication description, it was conjectured by
Entov and Yarin \cite{EYa84} and confirmed in \cite{CEFLM06,TLEAD18}
that the shape of the thinning thread could be described, at least for
sufficiently slow polymer relaxation, by a fluid with {\it infinite}
relaxation time. In that case, since stresses do not relax, one expects
the liquid to converge toward a stationary state, where surface tension
and elastic forces are balanced \cite{EYa84,MPFPP10,EF_book}. Our
work shows that, for example in the framework of an Oldroyd B fluid,
the limit $\lambda\rightarrow\infty$ corresponds exactly to minimizing
the sum of elastic and surface tension energy for a neo-Hookean
material. This has previously been shown only in the lubrication limit
\cite{EYa84,EF_book}. Within a lubrication description, it can be
shown that a stationary state will still be reached for a
Johnson-Segalman fluid, as long as the slip coefficient $a>1/2$.
However, below we will show that this result appears to be an artefact
of the lubrication (or slender-jet) approximation. Our full simulations
show that no stationary state is reached as soon as $a$ falls short
of the affine limit $a=1$.

To test these ideas, we simulate the Oldroyd B equations
\eqref{NS},\eqref{sigma_OB} in the limit of $\lambda\rightarrow\infty$,
taken such that $\mu = \eta/\lambda$ remains finite. Then the polymeric
stress is governed by $\overset{\triangledown}{\boldsymbol{\sigma}}_p=0$.
The collapse is driven by surface tension, and the stress
boundary condition at the free surface is
\beq
{\bf n}\cdot\boldsymbol{\sigma} = -\kappa{\bf n},
\label{stress_bc}
\eeq
where
\beq
\kappa = \frac{1}{h(1+h_{z}^2)^{1/2}} -
\frac{h_{zz}}{(1+h_{z}^{2})^{\frac{3}{2}}}, \quad
{\bf n} = \frac{{\bf e}_r-{\bf e}_z h_z}{(1+h_{z}^2)^{1/2}}
\label{kappa}
\eeq
are (twice) the mean curvature and the surface normal, respectively.
If $h(z,t)$ is the thread profile, the kinematic boundary condition becomes
\beq
\frac{\partial h}{\partial t} + u_z(z,h)\frac{\partial h}{\partial z} =
u_r(z,h),
\label{kin_bc}
\eeq
where ${\bf v} = u_r{\bf e}_r + u_z{\bf e}_z$ in cylindrical coordinates.
As an initial condition, we take the free surface shape
\beq
h(z,0) \equiv h_0(z) = R_0\left[1 -
  \epsilon\cos\left(\frac{z}{2R_0}\right)\right],
\label{h0}
\eeq
and both the velocity field and stresses vanish initially. Boundary
conditions are periodic.

We have carried out  a simulation for
a fluid cylinder of radius $R_0$, which is slightly perturbed according
to \eqref{h0} with $\epsilon=0.05$, $\eta_s/\sqrt{\rho\gamma R_0^3}=0.79$
and $\mu R_0/\gamma=0.0119$. In order to calculate the nonlinear time
evolution of the flow, we apply the boundary fitted coordinate method,
where the liquid domain  is mapped onto a rectangular domain through a
coordinate transformation. The hydrodynamic equations are discretized
in this domain using fourth-order finite differences, with 22 equally
spaced points in the radial direction and 1000  equally spaced points
in the axial direction. An implicit time advancement is performed using
second-order backward finite differences with a fixed time step
$0.05\sqrt{\rho R_0^3/\gamma}$; details of the numerical procedure
can be found elsewhere \cite{Herrada2016}.

\begin{figure}
\centering
\includegraphics[width=1\hsize]{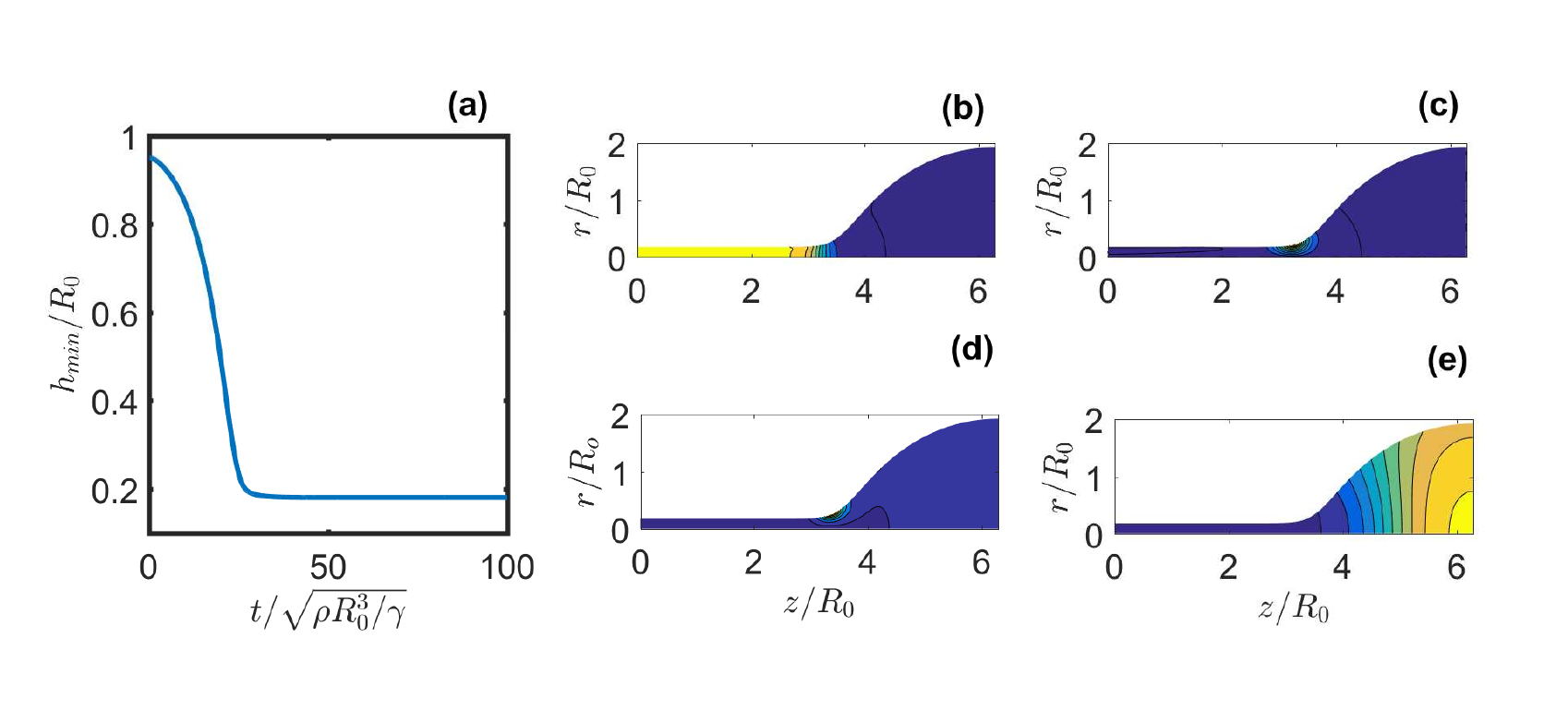}
\caption{(a) Time evolution of the minimum thread radius for the
viscoelastic fluid simulation  with $\epsilon=0.05$,
$\eta_s/\sqrt{\rho\gamma R_0^3}=0.79$ and $\mu R_0/\gamma=0.0119$.
Polymeric stresses in the final state of the fluid simulation:
(b) $\sigma_{p,zz}$;
(c) $\sigma_{p,zr}$; (d) $\sigma_{p,rr}$;
(e) $\sigma_{p,\theta\theta}.$
The purple to yellow color gradient grows from  minimum to maximum stress.
}
\label{fig:timeevolution}
\end{figure}

In Fig.~\ref{fig:timeevolution}-(a) we show the minimum thread radius,
$h_{min}$, as function of time.  As the bridge collapses, elastic
stress builds up until it is balanced by surface tension, as seen
in Fig.~\ref{fig:timeevolution}. At this point the solution becomes stationary,
time derivatives vanish, and the velocity goes to zero. As a
result, the solvent viscosity does not affect the final state, seen in
Fig.~\ref{fig:timeevolution}-(b)-(d), which shows the different components of the
stress tensor. The axial stress $\sigma_{p,zz}$ is highest inside the
thread, where fluid elements are stretched the most in the axial
direction. Radial stresses $\sigma_{p,rr}$, on the other hand, are
most pronounced inside the drop, where fluid elements are stretched in the
radial direction.

\subsection{The neo-Hookean solid}

\begin{figure}
\centering
\includegraphics[width=0.7\hsize]{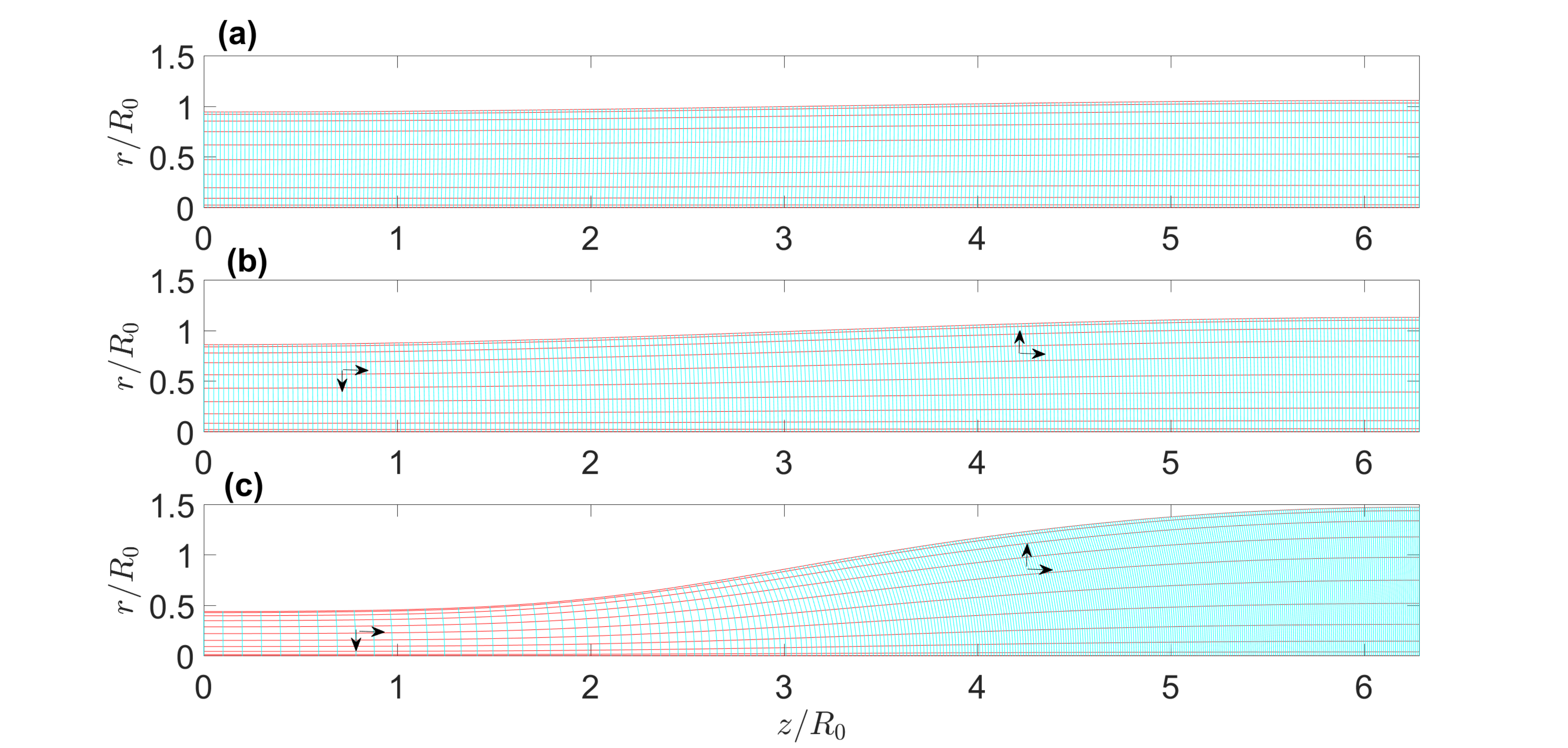}
\caption{The elastic simulation: (a) The reference state,
$\mu R_0/\gamma=\infty$; (b) $\mu R_0/\gamma=0.2$; (c) $\mu R_0/\gamma=0.1.$
Red (cyan) lines describe constant values of $\eta$ ($\xi$).  The arrows indicate the direction of the mesh deformation.
}
\label{fig:elastic}
\end{figure}

\begin{figure}
\centering
\includegraphics[width=0.7\hsize]{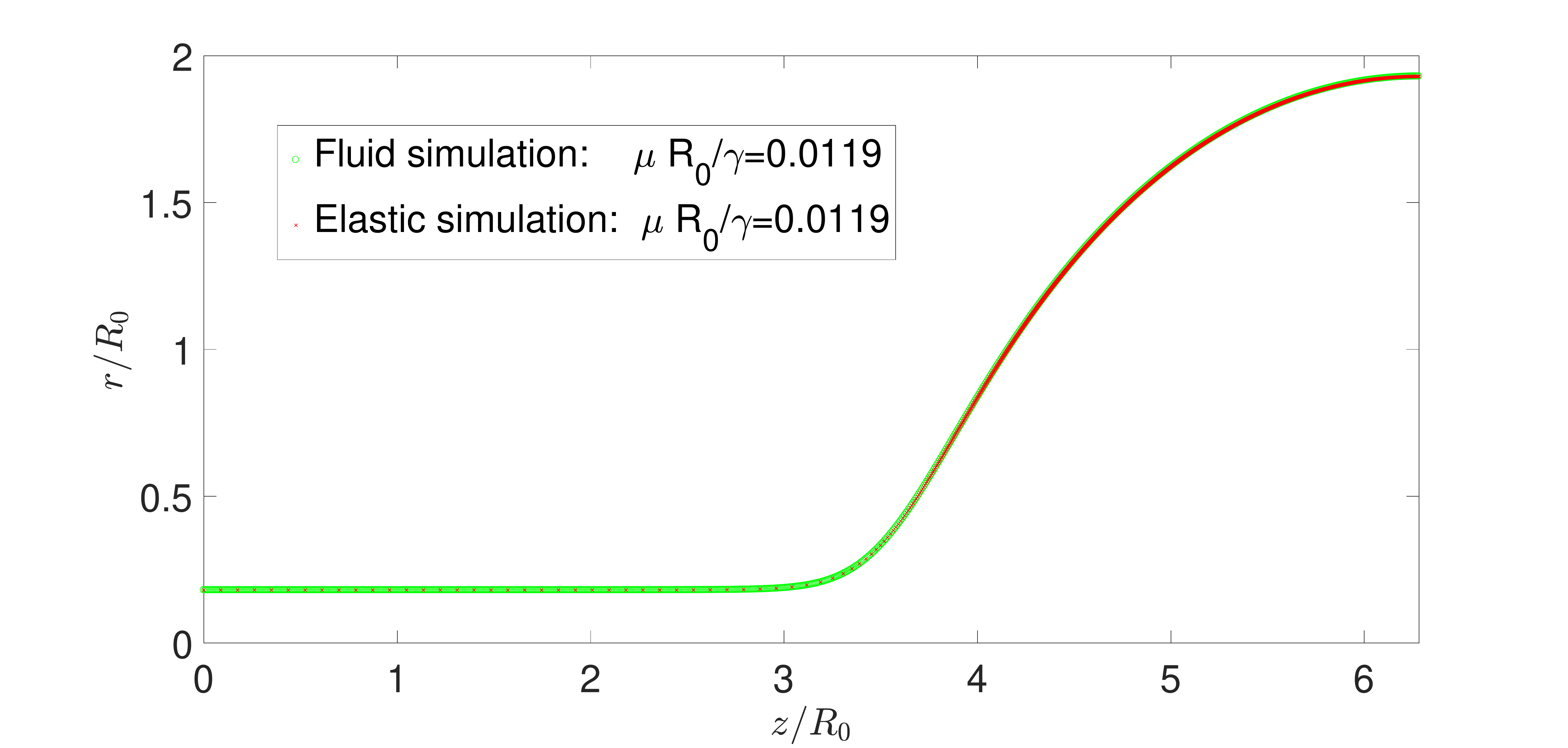}
\caption{A comparison between the final shape of an elastic bridge driven
by surface tension, using an Oldroyd B fluid with $\lambda=\infty$ (green) and
a neo-Hookean elastic material (red). The results are identical.
}
\label{fig:comp}
\end{figure}

We now calculate the steady state of an elastic neo-Hookean
material using non-linear elasticity, as described by
\beq
\label{eq:neohookstress}
\boldsymbol{\sigma} = \mu
\left( \mathbf F \cdot \mathbf F^T - \mathbf I \right) -p{\bf I}
\eeq
and subject to the incompressibility constraint $J=\det(\mathbf F)=1$.
The pressure is adjusted such that $J=1$ is satisfied. Instead of a
dynamical equation, the condition for equilibrium reads
$\nabla \cdot \boldsymbol{\sigma}=\mathbf 0$, i.e. (\ref{NS}) with
$\mathbf v=\mathbf 0$, with elasto-capillary boundary condition
(\ref{stress_bc}).

To determine the final state of the collapsed cylinder, we have to solve
a nonlinear set of equations corresponding to the above conditions,
based on a mapping ${\bf x} = {\bf x}({\bf X})$. To this end,
we write the mapping in cylindrical coordinates:
$r = r(R,Z)$, $z = z(R,Z)$. The coordinates $R$ and $Z$ are the radial
and axial coordinates of the cylinder in the reference state.
Using general formulae for ${\bf F}$ in cylindrical
coordinates~\cite{Negahban12} , incompressibility amounts to
\beq
\det{\bf F} = \frac{r}{R}\left(\frac{\partial r}{\partial R}
\frac{\partial z}{\partial Z}-\frac{\partial r}{\partial Z}
\frac{\partial z}{\partial R}\right)= 1,
\label{incompr}
\eeq
while the stress can be computed from the finger tensor
\beq
\mathbf B = \mathbf F \cdot \mathbf F^T =
\begin{pmatrix}
\left(\left(\frac{\partial r}{\partial R}\right)^2 +
\left(\frac{\partial r}{\partial Z}\right)^2\right)
& 0 & \left(\frac{\partial r}{\partial R}\frac{\partial z}{\partial R}
+\frac{\partial r}{\partial Z}\frac{\partial z}{\partial Z}\right)\\
0 & \left(\frac{r}{R}\right)^2 & 0  \\
\left(\frac{\partial r}{\partial R}\frac{\partial z}{\partial R}
+\frac{\partial r}{\partial Z}\frac{\partial z}{\partial Z}\right)
& 0 & \left(\left(\frac{\partial z}{\partial Z}\right)^2 +
\left(\frac{\partial z}{\partial R}\right)^2\right)
\end{pmatrix}.
\label{true}
\eeq
The solution depends on the dimensionless number
$R_0\mu/\gamma$. In the ``soft" limit where the thread becomes very thin
$r \ll R_0$, the thickness is of thread scales as the elasto-capillary
length scale
\cite{EF_book}
\beq
\ell_e = \left(\frac{\mu R_0^4}{\gamma}\right)^{1/3}.
\label{le}
\eeq

To solve the problem numerically, we define the reference state by
\[
R=h_0(\xi)\eta, \quad Z=\xi,
\]
with the elastic domain defined by $\eta \in [0,1]$ and
$\xi \in [0,2\pi R_0]$;
$h_0$ is once again defined by \eqref{h0}, and $\epsilon=0.05.$
We are looking for two unknown
functions $f$ and $g$, where $r=r(R,Z)=f(\eta,\xi)$ and
$z=z(R,Z)=g(\eta,\xi)$, as well as the pressure $p(\eta,\xi)$.
These three unknowns are found from solving the three equations
\eqref{incompr}, \eqref{NS} at steady state, and \eqref{stress_bc}.
The free surface $h(z)$ then is given by the parametric representation
$h(g(1,\xi)) = f_1(1,\xi)$, from which the curvature $\kappa$ can be
evaluated. 
The domain is discretized using fourth-order finite differences with 
301 equally spaced points in the $\xi$ direction and 
11 Chebyshev collocation points in the $\eta$ direction. 
For the results presented in Fig.~\ref{fig:comp}, a finer mesh was used 
with 2001 equally spaced points in the $\xi$ direction. 

The  resulting system of non linear
equations  is solved using a Newton-Raphson technique~\cite{Herrada2016}.
We solve  the problem by starting with  the reference state as initial
guess  and $\mu R_0/\gamma$ sufficiently large ($\mu R_0/\gamma=100$)
to ensure the convergence of the  Newton-Raphson iterations. Once we
get a solution, we use this solution in a new run with a smaller value
of $\mu R_0/\gamma$.

The result is shown in Fig.~\ref{fig:elastic}, which describes the
deformation of the mesh as well as of the free surface, for various values
of $\mu R_0/\gamma$. The resulting shapes closely resemble those in
Fig.~\ref{fig:intro} at moderate stiffness, and agree with simulations
in \cite{XB17}. In Fig.~\ref{fig:comp} we compare the elastic equilibrium
state (red dots) with the stationary state reached in the simulation
of the Oldroyd-B model for $\mu R_0/\gamma=0.0119$.
The agreement is perfect -- illustrating that Oldroyd-B converges to
neo-Hookean in the limit of $\lambda \rightarrow \infty$, as we
have shown in the present paper. A detailed similarity analysis of
this problem is provided in \cite{EHS19}.

\subsection{The Johnson-Segalman fluid}

\begin{figure}
\centering
\includegraphics[width=0.5\hsize]{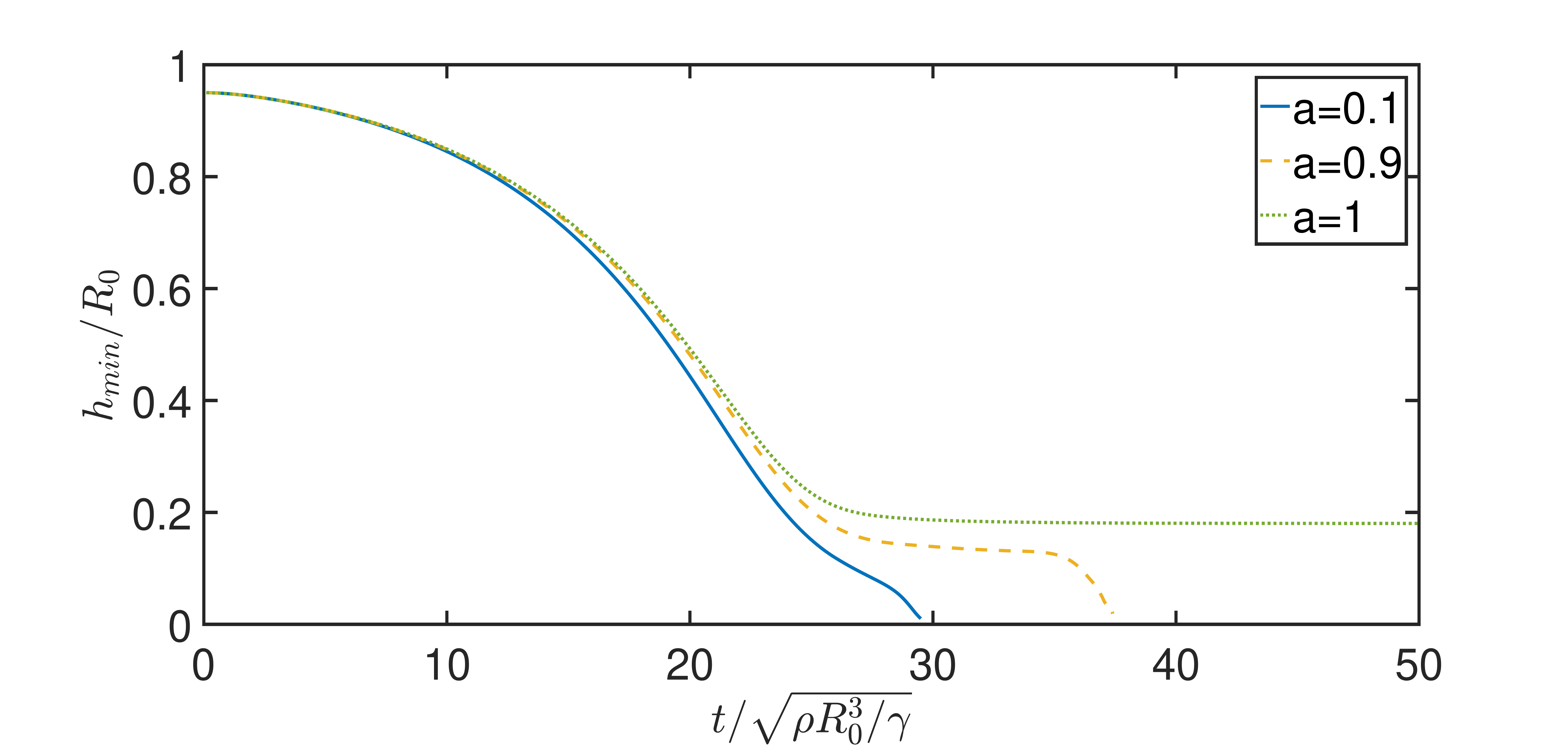}
\caption{The minimum thread radius $h_{min}$ for a Johnson-Segalman
fluid, as a function of time for for different values of $a$. Only in
the case of non-affine motion ($a=1$), the thread attains a
static elastic solution. Breakup is observed for all values $a\neq 1$.
}
\label{fig:timeevolution1}
\end{figure}

\begin{figure}
\centering
\includegraphics[width=0.7\hsize]{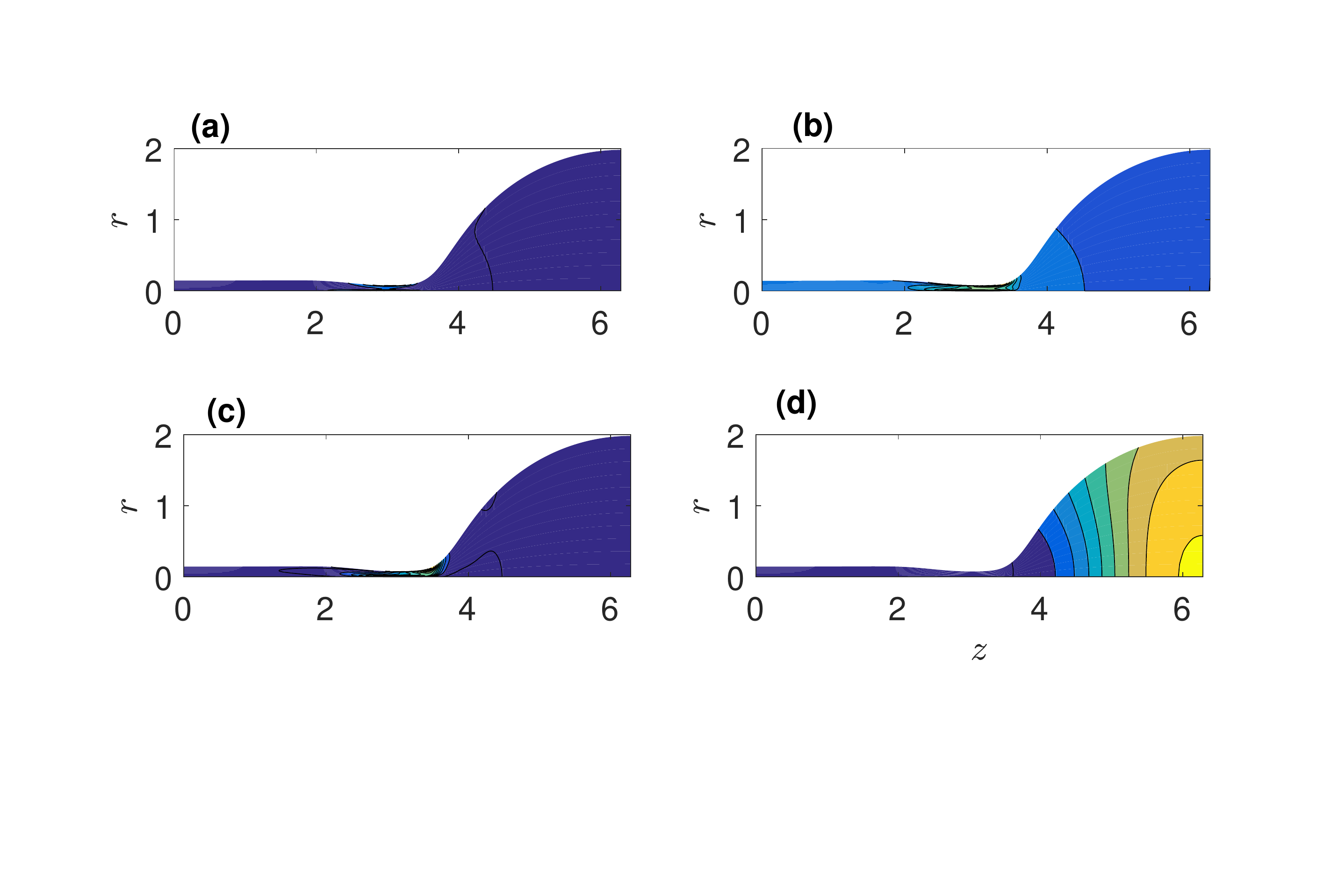}
\caption{Polymeric stresses at a time close to the breakup of the
Johnson-Segalman fluid simulation with $a=0.9$: (a) $\sigma_{p,zz}$;
(b) $\sigma_{p,zr}$; (c) $\sigma_{p,rr}$; (d) $\sigma_{p,\theta\theta}$.
The purple to yellow colour gradient grows from  minimum to maximum stress.
}
\label{fig:JS}
\end{figure}

The above comparison was an illustration of our result which
assigns a unique elastic limit to the Oldroyd B model for large
relaxation times. Now we consider a Johnson-Segalman fluid with
infinite relaxation time, characterized by
$\overset{\triangledown}{\left(\mathbf A\right)}_a = 0$.
The other equations remain the same. Previous analysis of
the long-wavelength limit has shown \cite{EF_book} that there
can be no surface tension - elastic balance for $a<1/2$. This result
was obtained through an analysis of a thread of uniform thickness.
However, since the true solution is not a uniform thread throughout,
$a>1/2$ is only a necessary condition for a stationary state.

Johnson-Segalman fluid simulations are carried out by integrating
\eqref{eq:JSscaled} with $\lambda=\infty$  for different values of
$a$ with  the same numerical technique described at the beginning
of the section, using the same parameters $\epsilon=0.05$,
$\eta_s/\sqrt{\rho\gamma R_0^3}=0.79$, and $\mu R_0/\gamma=0.0119$ as
before. Figure~\ref{fig:timeevolution1} shows $h_{min}$ as function of
time for three different values of the slip parameter $a$. As it
can be seen in the figure, the solution reaches a steady state only
for the affine case $a=1$ while for the other cases the numerical
solution breaks. Particular interesting is the case, $a = 0.9$,
which according to the lubrication analysis should reach a
stationary state. However, as seen in Fig.~\ref{fig:JS},
the solution is qualitatively different from the case $a=1$:
the thread becomes non-uniform in space and continues to evolve
in time until the solution breaks.

\subsection{A mixed Eulerian-Lagrangian method for
  neo-Hookean solid and Newtonian fluids}
\begin{figure}
\centering
\includegraphics[width=0.6\hsize]{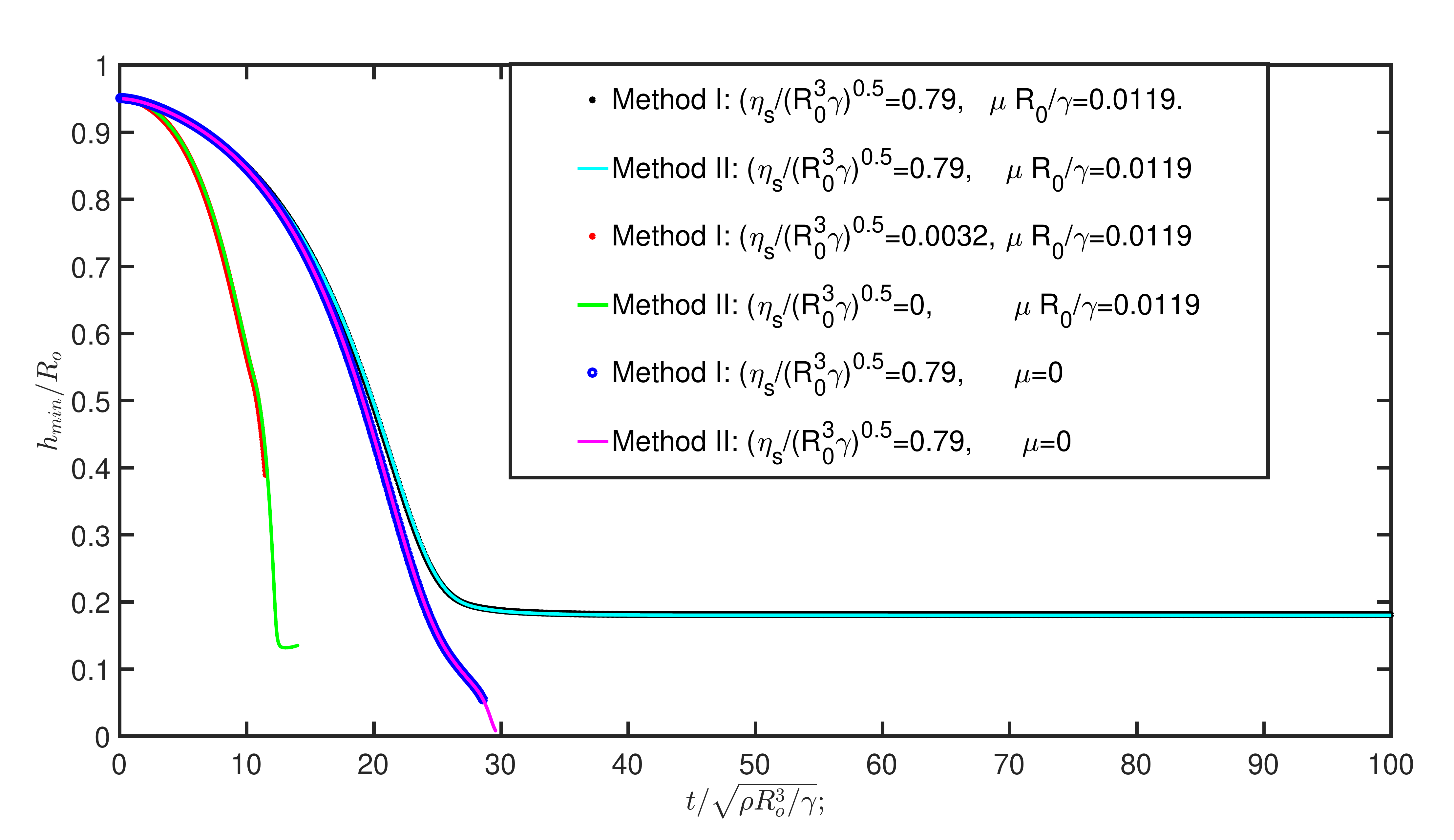}
\caption{A comparison of the Eulerian method based on the Oldroyd-B
equations (method I), and the mixed Eulerian-Lagrangian method based 
on finding the Lagrangean map (method II) for the collapse of a liquid
bridge. The time evolution of the minimum thread radius is shown for 
three different materials: (i) an elastic liquid with solvent (black, cyan),
(ii) a purely elastic liquid without solvent (red, green), and (iii) a
Newtonian liquid (blue, magenta). }
\label{fig:timeevolution2}
\end{figure}

In the above, we have compared numerical simulations to investigate
the elastic limit $\lambda\rightarrow\infty$. On one hand, we have
employed purely Eulerian ideas using the Oldroyd-B equations, as
shown in Fig.~\ref{fig:timeevolution}, to advance the state of the
system in time. Eventually, a steady state is reached, and since
the velocity goes to zero, the solvent does not contribute to the
stress, and the steady state represents a purely elastic balance.
However, taking $\eta_s$ to zero represents a very singular
limit, since the velocity disappears from the momentum equation
\eqref{NS}, apart from inertial terms on the left, which are very small.

On the other hand, we used the purely Lagrangian description
\eqref{eq:neohookstress}, where the Finger tensor ${\bf B}$ is given
in terms of the Lagrangian map ${\bf x}({\bf X},t)$. By solving the
nonlinear equation $\grad \cdot \boldsymbol{\sigma}=\mathbf 0$,
we can find the equilibrium state, which is in perfect agreement with
the long-time limit of the Eulerian simulation, as seen in
Fig.~\ref{fig:comp}.

By using the transformations between Eulerian and Lagrangian
formulations layed out in Section~\ref{sub:kinematics}, we can
construct a much more versatile method, which combines the advantages
of both. Instead of Eulerian fields, everything is solved for the
mapping ${\bf x}({\bf X},t)$. Calculating derivatives, we obtain
the deformation gradient tensor ${\bf F}$, as well as $\partial {\bf X}/\partial t$,
from which we find the Eulerian velocity ${\bf v}$ via \eqref{velocity}.
Instead of the incompressibility condition $\grad\cdot{\bf v} = 0$,
we implement $J = \det({\bf F}) = 1$. if we were to deal with a
compressible liquid, we could use \eqref{copressible_L} instead of
the continuity equation. 
  
To illustrate that we are able to go all the way from an elastic
solid to a pure viscous liquid using the same method, we consider
the total stress \eqref{sigma_OB} in the limit
$\lambda\rightarrow\infty$, so that we can write 
\beq
\label{eq:neohookstress1}
\boldsymbol{\sigma} =
\mu\left( \mathbf F \cdot \mathbf F^T - \mathbf I \right)
+\eta_s \boldsymbol{\dot{\gamma}} - p{\bf I}. 
\eeq
For $\mu>0$ and $\eta_s=0$, \eqref{eq:neohookstress1} corresponds to a
neo-Hookean solid, while if  $\mu=0$ and $ \eta_s>0$ the material is
a pure Newtonian fluid.

In order to solve the problem numerically, we need to be able to
compute partial derivatives of in both the current state and in the
reference state. To achieve that, a numerical reference domain
$\eta \in [0,1]$ and $\xi \in [0,2\pi R_0]$ is introduced. The current and
reference state are both mapped to the numerical domain through
\[
R=R_n(\xi,\eta,t),\quad Z=Z_n(\xi,\eta,t);\quad r=r_n(\xi,\eta,t)\quad
z=z_n(\xi,\eta,t)
\]
at each point in time. To close the system of equations for these
additional unknowns, we have chosen to set
\beq
R_n=h_0(Z_n)\eta, \quad z_n=\xi,
\label{mapping1}
\eeq
where  $h_0$ is once again defined by \eqref{h0}, and $\epsilon=0.05.$
This choice allows us to keep a steady distribution of points in the
axial direction in  the current state.

In summary, we are looking for the unknown functions
$r_n(\xi,\eta,t)$ ,$z_n(\xi,\eta,t)$, $R_n(\xi,\eta,t)$,
$Z_n(\xi,\eta,t)$ $v_z(\xi,\eta,t)$,  $v_r(\xi,\eta,t)$, and
$p(\xi,\eta,t)$. These seven unknowns are found from solving the
momentum equation \eqref{NS} with stress \eqref{eq:neohookstress1},
the incompressibility constraint \eqref{incompr}, the velocity 
\eqref{velocity}, the mapping \eqref{mapping1}, and the free surface
condition \eqref{stress_bc}. 
Then the free surface $h(z,t)$ is given by the parametric representation
$h(z_n(\xi,\eta,t),t) = r_n(\xi,1,t)$, from which the curvature
$\kappa$ can be evaluated. For the simulations reported in
Fig.~\ref{fig:timeevolution2}, the domain is discretized using fourth-order
finite differences with 801 equally spaced points in the $\xi$ direction and
11 Chebyshev collocation points in the $\eta$ direction. An implicit time
step is performed using second-order backward finite differences with a
fixed time step $0.01\sqrt{\rho R_0^3/\gamma}$. The  resulting system of
nonlinear equations is solved using a Newton-Raphson
technique~\cite{Herrada2016}.

In Fig.~\ref{fig:timeevolution2} we show the minimum thread radius,
$h_{\rm min}$, as function of time for three different materials using the
mixed Eulerian-Lagrangian approach (method II, solid lines). These
results are compared to our earlier Eulerian approach, using the Oldroyd-B
equations (method I, symbols). The results are virtually identical, but the
method II is often found to be more stable, in particular at small
values of the solvent viscosity, since the Eulerian method becomes
singular in that limit.

The black dots (method I) and solid cyan line (method II) corresponds
to the case considered in the previous section, for which $\eta_s>0$, 
and for which the system evolves toward a steady state. This is seen in 
Fig.~\ref{fig:timeevolution2} by the fact that $h_{\rm min}$ approaches
and then stays at a constant value, representing a elastic capillary
balance. The black dots (method I) are the same data as seen before in
Fig.~\ref{fig:timeevolution}, but the cyan line now shows the complete
time evolution toward the steady state, using the Lagrangian map;
the two results are indistinguishable from one another.

The red dots
(method I) are for the same system, but with a very small solvent
viscosity. 
However, the method fails even before coming close to
a stationary state, owing to the singular nature of the small-$\eta_s$
limit. The solid green line (method II) corresponds to a pure
neo-Hookean solid, and can now been treated owing to the mixed
Euler-Lagrangian method. The initial dynamics represent a balance
between elasticity and inertia, and is much more rapid since in this
case there is not dissipation ($\eta_s=0$). Up to the point where
the Eulerian method fails, there is close agreement between both methods. 
As elastic stress builds up, the inertial motion is arrested
suddenly, and the bridge rebounds. This deforms the interface so much
that the mapping \eqref{mapping1}) becomes singular, and the numerics
break down, because the shape can no longer be represented properly. 

Finally, putting $\mu = 0$ and letting $\eta_s$ remain finite,
we obtain a pure Newtonian fluid, in which there are no more elastic
stresses. Both methods (blue dots, method I, solid magenta curve,
method II), deliver the same result, in which there are no elastic
stresses to balance surface tension. As a result, the solution pinches
off in finite time, with the minimum radius $h_{\rm min}$ a linear
function of time \cite{EV08}. In our numerical examples we have
only considered cases with infinite relaxation time, but the
Eulerian-Lagrangian scheme would work equally well for a finite $\lambda$, 
which interpolates smoothly between fluid and elastic behavior. 

\section{Discussion}
\label{sec:discussion}

In summary, we have provided a detailed overview of how viscoelastic
models are related to the theory of elasticity. By analyzing the
kinematics in the limit of large relaxation times, we have identified
a systematic route to express the energy balance in viscoelastic flows.
This is based on the separation of the reversible elastic energy from
the dissipation associated to relaxation phenomena. Our observations are
illustrated by a detailed analysis of the capillary instability of a
cylindrical jet, analyzed for both elastic and viscoelastic materials.
An important observation is that the presence of non-affine motion in polymer
solutions has a dramatic effect, as it will lead to pinching of the thread
even in the limit of infinite relaxation times. This illustrates that
non-affine viscoelastic liquids do not have any counterpart within the
theory of elasticity.

From a general viewpoint, the explicit connection between elasticity
and viscoelasticity may provide an original perspective to problems
in either fields. For example, in \cite{EHS19} we have exploited the elastic 
correspondence to resolve the breakup of viscoelastic threads, 
which are traditionally studied e.g. by Oldroyd-B or FENE-P models. 
From a numerical perspective, the analysis developed here 
provides a new approach towards computational challenges. 
For example, the neo-Hookean simulation for elastic threads turned out
to be very efficient, and we have demonstrated how such schemes can 
also be extended to Newtonian fluids. Conversely, we note that using
viscoelastic liquids with infinite relaxation time could offer an
attractive, fully Eulerian approach to fluid-structure interaction problems.

\section*{Acknowledgments}
We gratefully acknowledge Anthony Beris and Alexander Morozov for providing 
detailed feedback on the manuscript. 
J. E. acknowledges the support of Leverhulme Trust International Academic
Fellowship IAF-2017-010, and is grateful to Howard Stone and his group
for welcoming him during the academic year 2017-2018. 
M. A. H. thanks the Ministerio de Econom\'{\i}a y competitividad for
partial support under the Project No. DPI2016-78887-C3-1-R
J.H.S. acknowledges support from NWO through VICI Grant No. 680-47-632, 
and A.P. from European Research Council (ERC) Consolidator Grant No. 616918.

\appendix

\section{Rheological models}
\label{app:rheo}

Here we list a number of frequently considered rheological models, and
show how they fit into our formalism, which consists in specifying an
elastic energy $W({\bf A})$ as well as a relaxation equation
$\lambda\overset{\triangledown}{\bf A} = f({\bf A},\dot{\boldsymbol{\gamma}})$.
Then the polymeric stress as well as the dissipation can be calculated
from \eqref{sigma_A} and \eqref{P} in the affine case, and
\eqref{sigma_P_Hooke} for the non-affine case.

\begin{enumerate}

\item {\bf Oldroyd B / Upper convected Maxwell model} \\

As described in Subsection~\ref{sub:ex}, in that case the equations can be
written
\beq
\boldsymbol{\sigma}_p = \mu\left( \mathbf A - \mathbf I \right), \quad
\overset{\triangledown}{{\mathbf{A}}}=
-\frac{1}{\lambda}\left( \mathbf A - \mathbf I \right);
\label{eq_ucmm}
\eeq
the elastic energy and dissipation are
\beq
W=\frac{\mu}{2}\left({\rm tr}({\bf A})-3\right), \quad
\epsilon_p =  \frac{W}{\lambda}.
\label{WP_ucmm}
\eeq

According to \eqref{sigma_OB}, the deviatoric stress $\boldsymbol{\tau}$
is the sum of solvent and polymer contributions. In the Oldroyd B model,
both can be combined into the single equation
\beq
\boldsymbol{\tau} + \lambda\overset{\triangledown}{{\boldsymbol{\tau}}} =
\eta\dot{\boldsymbol{\gamma}} + \lambda\eta_s
\overset{\triangledown}{\dot{\boldsymbol{\gamma}}}.
\label{OB_single}
\eeq
In the limit of vanishing shear rate, \eqref{OB_single}
describes a Newtonian fluid of total viscosity $\eta = \eta_s + \eta_p$,
the sum of polymeric and solvent contributions.

\item {\bf Oldroyd A / Lower convected Maxwell model}\\

As described in Subsection~\ref{sub:non-affine},
\beq
\boldsymbol{\sigma}_p = -\mu\left( \mathbf A - \mathbf I \right), \quad
\overset{\vartriangle}{\bf A} =
-\frac{1}{\lambda}\left( \mathbf A - \mathbf I \right),
\label{lc_Maxwell2}
\eeq
describes the evolution, and
\beq
W=\frac{\mu}{2}\left({\rm tr}({\bf A})-3\right), \quad
\epsilon_p =  \frac{W}{\lambda}.
\label{WP_lcmm}
\eeq
are elastic energy and dissipation, respectively.

\item {\bf FENE-P model}\\

As discussed in Subsection~\ref{sub:ex}, with
$I_1 = {\rm tr}(\mathbf A)$ and $f(I_1)= \frac{L^2-3}{L^2-I_1}$,
\beq
\boldsymbol{\sigma}_p = \mu f(I_1)\left( \mathbf A - \mathbf I \right), \quad
\overset{\triangledown}{{\mathbf{A}}}=
-\frac{1}{\lambda}\left(f(I_1)\mathbf A - \mathbf I \right).
\label{eq_FENE-P}
\eeq
so that
\beq
W=\frac{\mu}{2}(L^2-3)\ln\left(f(I_1)\right), \quad
\epsilon_p = \frac{\mu}{2\lambda}f(I_1)\left(I_1f(I_1)-3\right),
\label{WP_FENE-P}
\eeq
using \eqref{eq:p_simp}.

\item {\bf Giesekus model}\\

This is a phenomenological model \cite{BAH87,Larson99} which introduces
a term quadratic in $\sigma_p$ into the equation of motion, which
also limits the maximum value of the stress; however, the stress may
become arbitrarily large for sufficiently strong flow:
\beq
\boldsymbol{\sigma}_p +
\lambda \overset{\triangledown}{\boldsymbol{\sigma}}_p +
\alpha\frac{\lambda}{\eta_p}{\boldsymbol{\sigma}}_p \cdot {\boldsymbol{\sigma}}_p
= \eta_p \boldsymbol{\dot{\gamma}}.
\label{Giesekus}
\eeq
This can be written as
\beq
\boldsymbol{\sigma}_p = \mu\left( \mathbf A - \mathbf I \right), \quad
\overset{\triangledown}{{\mathbf{A}}}=
-\frac{1}{\lambda}\left[{\bf A} - {\bf I} + \alpha
\left({\bf A} - {\bf I}\right)^2 \right],
\label{eq_Giesekus}
\eeq
so that the elastic energy is once more neo-Hookean, and the dissipation is:
\beq
W=\frac{\mu}{2}\left(I_1-3\right), \quad
\epsilon_p = \frac{\mu}{2\lambda}\left(I_1-3\right) +
\frac{\mu\alpha}{2\lambda}\left({\bf A}:{\bf A}-2{\rm tr}({\bf A})+3\right).
\label{WP_Giesekus}
\eeq

\item {\bf Johnson-Segalman model}\\

As laid out in Subsection~\ref{sub:non-affine}, the Johnson-Segalman
model is written
\beq
\boldsymbol{\sigma}_p = a\mu\left( \mathbf A - \mathbf I \right), \quad
(\overset{\triangledown}{{\mathbf{A}}})_a=
-\frac{1}{\lambda}\left( \mathbf A - \mathbf I \right).
\label{eq_JS}
\eeq
The energy is once more neo-Hookean, and the dissipation is
according to \eqref{non-affine_diss}:
\beq
W=\frac{\mu}{2}\left(I_1-3\right), \quad
\epsilon_p = \frac{\mu}{2\lambda}\left(I_1-3\right) = \frac{W}{\lambda}.
\label{WP_JS}
\eeq

\end{enumerate}

\section{Curvilinear formulation of viscoelasticity}
\label{app:curvi}

\subsection{Kinematics of deformation}

The purpose of this appendix is to rephrase the results of the main
text in terms of curvilinear coordinates. This allows for a rigorous
analysis of the physical assumptions underlying the equation of motion
for the conformation tensor, $\mathbf{A}$. For a detailed description of
kinematics discussed below, the reader can refer to the book by
Green \& Zerna~\cite{green2002}.

We thus define curvilinear coordinates $q^i$, which are material points that move along affinely with the flow. The current position vector is defined as $\mathbf x(q^i,t)$, while the position of the reference (or initial) configuration reads $\mathbf X(q^i)=\mathbf x(q^i,t=0)$. The latter is independent of time. The distance $ds$ between two neighboring points $q^i$ and $q^i+dq^i$ reads

\begin{equation}\label{eq:ds}
ds^2 = \left(\frac{\partial \mathbf x}{\partial q^i} \cdot   \frac{\partial \mathbf x}{\partial q^j} \right) dq^i dq^j = g_{ij}dq^i dq^j,
\end{equation}
where $g_{ij}$ is the current metric tensor.
Similarly, the reference distance $dS$ follows as

\begin{equation}\label{eq:dsref}
dS^2 = \left(\frac{\partial \mathbf X}{\partial q^i} \cdot   \frac{\partial \mathbf X}{\partial q^j} \right) dq^i dq^j = G_{ij}dq^i dq^j,
\end{equation}
where $G_{ij}$ is the reference metric. Stretching of material elements follows from changes in length

\begin{equation}
ds^2 - dS^2 = \left(g_{ij} - G_{ij} \right) dq^i dq^j,
\end{equation}
so strain is encoded in the difference between current and reference metric.

We now construct the current vector space, using the covariant and contravariant basis vectors

\begin{equation}\label{eq:basis}
\mathbf e_i = \frac{\partial \mathbf x}{\partial q^i}, \quad \mathbf e^j = \frac{\partial q^j}{\partial \mathbf x},
\end{equation}
derived from the current position $\mathbf x$. The covariant basis vectors $\mathbf{e}_i$ are local tangents to the material lines in the deformed configuration. The contravariant vectors $\mathbf{e}^i$ form a reciprocal basis, owing to the property $\mathbf e_i \cdot \mathbf e^{j} =dq^j/dq^i = \delta_i^j$.
Using this basis, we can define the metrics
\beq
g_{ij} = \mathbf e_i \cdot \mathbf e_j, \quad
g^{ij}=\mathbf e^i \cdot \mathbf e^j,
\label{g_metric}
\eeq
so that $g_{ij}$ can be used to lower indices, while the inverse
$g^{ij}$ raises indices. Similarly, one can construct the reference
vector space, using ``reference" basis vectors
\begin{equation}\label{eq:refbasis}
\mathbf E_i = \frac{\partial \mathbf X}{\partial q^i},\quad \mathbf E^j = \frac{\partial q^j}{\partial \mathbf X}.
\end{equation}
The associated metric for this basis, as well as its inverse,
are defined by
\beq
G_{ij}=\mathbf E_i \cdot \mathbf E_j, \quad
G^{ij}=\mathbf E^i \cdot \mathbf E^j;
\label{Gc_metric}
\eeq
the $\mathbf{E}_i$ are local
tangents to the material lines in the reference configuration.

We now wish to express the metrics in terms of the mapping $\mathbf F =\partial \mathbf x/\partial \mathbf X$. In particular, we wish to show that Green's deformation tensor $\mathbf C = \mathbf F^T \cdot \mathbf F$ and the Finger tensor $\mathbf B=\mathbf F \cdot \mathbf F^T$ can be written as

\begin{equation}\label{eq:cb}
\mathbf C = g_{ij} \mathbf E^i \otimes \mathbf E^j, \quad \mathbf B = G^{ij} \mathbf e_i \otimes \mathbf e_j.
\end{equation}
To demonstrate this, we write  (\ref{eq:ds}) as

\begin{equation}
ds^2 =\left( \frac{\partial \mathbf x}{\partial \mathbf X} \cdot \frac{\partial \mathbf X}{\partial q^i} \right)^T
\cdot \left( \frac{\partial \mathbf x}{\partial \mathbf X} \cdot \frac{\partial \mathbf X}{\partial q^j} \right) dq^i dq^j
= \mathbf E_i \cdot \left( \mathbf F^T \cdot \mathbf F\right) \cdot \mathbf E_j \, dq^i dq^j.
\end{equation}
Comparing to (\ref{eq:ds}), we indeed see that $g_{ij}$ are the covariant components of $\mathbf C = \mathbf F^T \cdot \mathbf F$ when expressed using the basis $\mathbf E^i$. Hence, we obtain the first identity in (\ref{eq:cb}). It is important to keep track of the basis used to express the tensor~\cite{Hanna19}; for example, pairing $g_{ij}$ with the basis $\mathbf e^i$, one recovers the identity tensor, $\mathbf{I}=g_{ij}\mathbf{e}^i\otimes\mathbf{e}^j$.

Similarly we rewrite (\ref{eq:dsref}) as

\begin{equation} \label{eq:fingerinverse}
dS^2 =\left( \frac{\partial \mathbf X}{\partial \mathbf x} \cdot \frac{\partial \mathbf x}{\partial q^i} \right)^T
\cdot \left( \frac{\partial \mathbf X}{\partial \mathbf x} \cdot \frac{\partial \mathbf x}{\partial q^j} \right) dq^i dq^j
= \mathbf e_i \cdot \left( \mathbf F^{-T} \cdot \mathbf F^{-1} \right) \cdot \mathbf e_j \, dq^i dq^j.
\end{equation}
Now we see that $G_{ij}$ are the covariant components of $\mathbf B^{-1}=\mathbf F^{-T} \cdot \mathbf F^{-1}$ when expressed using the basis $\mathbf e^i$.  Since the inverse of the reference metric is defined as $G_{ik}G^{kj} = \delta^j_i$, we obtain the second identity in (\ref{eq:cb}). Again, pairing $G_{ij}$ with the basis $\mathbf E^i$, one recovers the identity tensor, $\mathbf{I}=G_{ij}\mathbf{E}^i\otimes\mathbf{E}^j$.

\subsection{Flow}

Now, we investigate the effect of flow on the metric. First, we define the velocity as

\begin{equation}
\mathbf v = \left(\frac{d\mathbf x}{dt}\right)_{q^i} = v^i \mathbf e_i,
\end{equation}
expressed using the basis defined by (\ref{eq:basis}), where from now
on $d/dt$ means time-derivative at constant material points $q^i$.
Using \eqref{g_metric} and \eqref{eq:basis}, the time-derivative of
the metric tensor is
\begin{eqnarray}
\frac{dg_{ij}}{dt} &=&  \left(\frac{\partial  \left(v^k \mathbf e_k\right)}{\partial q^i} \cdot   \mathbf e_j \right)  +
\left(\mathbf e_i \cdot \frac{\partial  \left(v^m \mathbf e_m\right)}{\partial q^j} \right)
\equiv  \left( v^k_{;i} \mathbf e_k \right) \cdot   \mathbf e_j   +  \mathbf e_i \cdot \left( \mathbf e_m \right) v^m_{;j} \nonumber \\
&=& g_{kj} v^k_{;j} + g_{mi} v^m_{;j}  = v_{j;i} + v_{i;j}  \equiv \dot{\gamma}_{ij},
\end{eqnarray}
where we used the definition of the covariant derivative, denoted
by $(..)_{;j}$. Hence, the rate of strain tensor
$\dot{\boldsymbol{\gamma}}$ directly gives the change of the metric
of material coordinates by the flow.
Remembering that $B^{ij}=G^{ij}$ (cf. \eqref{eq:cb}), we see that
the Finger tensor evolves according to
\begin{equation}\label{eq:up}
\frac{dB^{ij}}{dt}= \frac{d}{dt}\left( \frac{\partial q^i}{\partial \mathbf X} \cdot \frac{\partial q^j}{\partial \mathbf X} \right) = 0
\end{equation}
during flow. This time-derivative vanishes because the reference
state $\mathbf X(q^i)$ is independent of time. Note, however, that
the covariant components $B_{ij}$ are not constant in time, since
\begin{equation}\label{eq:low}
\frac{dB_{ij}}{dt}=\frac{d}{dt}\left(g_{ik} g_{jm} B^{km} \right) = 2 \dot{\gamma}_{ik} B^k_j.
\end{equation}

For  a general tensor $\mathbf A$, the derivatives $dA^{ij}/dt$ and $dA_{ij}/dt$ respectively correspond to the components of the upper and lower convected derivatives~\cite{O50,aris1990}, i.e.


\begin{equation}\label{eq:uplowcovariant}
\overset{\triangledown}{{\mathbf{A}}} = \frac{dA^{ij}}{dt} \mathbf e_i \otimes \mathbf e_j, \quad \quad
\overset{\vartriangle}{{\mathbf{A}}}  = \frac{dA_{ij}}{dt} \mathbf e^i \otimes \mathbf e^j.
\end{equation}
The equivalence with the definitions \eqref{up_relation} and
\eqref{down_relation} follow from transforming $A^{ij}$ respectively
$A_{ij}$ from the Lagrangian (co-moving) material coordinates $q^i$,
to an Eulerian coordinate system $\bar q^i$ that is fixed in space.
In this fixed coordinate system, the tensor components indicated
$\bar{A}^{ij}$ can be obtained using the transformation
$F^k_{i}=\partial \bar{q}^k/\partial q^i$.
Transforming $dA^{ij}/dt$ to the fixed frame then gives

\begin{eqnarray}
F^p_i \left( \frac{dA^{ij}}{dt} \right) F^q_j  &=& F^p_i \frac{d}{dt} \left[ (F^{-1})^i_k  \bar{A}^{km} (F^{-1})^j_m \right] F^q_j.
\label{A_t}
\end{eqnarray}
On the right hand side we recognize the definition \eqref{up_relation}
for $\overset{\triangledown}{{\mathbf{A}}}$, now in the form of the
components of the fixed Eulerian system $\bar q^i$.
In similar fashion
one derives (\ref{eq:uplowcovariant}) for the lower convected derivative.
Hence, $dB^{ij}/dt=0$ implies that the upper convected derivative of
the Finger tensor vanishes, i.e. $\overset{\triangledown}{{\mathbf{B}}}=0$.

\subsection{Elasticity}

In the theory of elasticity, the energy density $W$ is a function
of the invariants of $\mathbf B$. Assuming incompressibility
$I_3 = \det (\mathbf B)=1$, the energy is of the form $W(I_1,I_2)$,
where the first and second invariants are defined as

\begin{equation}
I_1 = B^i_i = g_{ij}B^{ij}, \quad
I_2  = \frac{1}{2}\left(\left[ g_{ij}B^{ij}\right]^2 - B_{ij} B^{ij} \right) = \frac{1}{2}\left( \left[ g_{ij}B^{ij}\right]^2 - g_{im} g_{jn} B^{mn} B^{ij} \right).
\end{equation}
The expression for stress is obtained from time-derivatives, according to the virtual work principle,

\begin{equation}\label{eq:virtual}
\frac{1}{2}\sigma^{ij} \dot{\gamma}_{ij} = \frac{dW}{dt} = W_1 \frac{dI_1}{dt}  + W_2 \frac{dI_2}{dt}.
\end{equation}
Using $dB^{ij}/dt=0$, we find

\begin{equation}
\frac{dI_1}{dt} = \frac{dg_{ij}}{dt} B^{ij} = \dot{\gamma}_{ij}B^{ij}, \quad
\frac{dI_2}{dt} = \frac{dg_{ij}}{dt} \frac{\partial I_2}{\partial g_{ij}}
= \dot{\gamma}_{ij} \left(I_1B^{ij} - B^i_n B^{nj} \right),
\end{equation}
both of which are proportional to $\dot{\gamma}_{ij}$. Hence, from (\ref{eq:virtual}) we can read off the stress as

\begin{equation}
\sigma^{ij}  = 2\frac{\partial W}{\partial g_{ij}} = 2 W_{1} B^{ij} + 2W_{2}\left( I_1  B^{ij}  - B^i_n B^{nj}  \right).
\end{equation}

\subsection{Viscoelasticity}

When an elastic \emph{liquid} is suddenly arrested, the polymer will relax towards an isotropic equilibrium confirmation. The system slowly forgets about the history of deformation prior to the arrest, and all stress and polymer stretches are ultimately relaxed. When expressing the polymer deformation in terms of the elementary lengths between $q^i$ and $q^i+dq^i$, we can still write

\begin{equation}
ds^2 - dS^2 = \left(g_{ij} - G_{ij} \right) dq^i dq^j,
\end{equation}
and we associate an elastic energy to the polymer strain. Owing to the
fading memory of the initial state $\mathbf x(q^i,t=0)$, however, the
object $G_{ij}$ can no longer be identified with the time-independent
metric of this initial condition. Instead, $G_{ij}$ reflects the metric
of the ``instantaneous reference state" $\mathbf X(q^i,t)$ that
progressively tends to evolve towards the current state. Hence, using
\eqref{Gc_metric} and \eqref{eq:refbasis}, we obtain
\begin{equation}
G_{ij} = \frac{\partial \mathbf X}{\partial q^i} \cdot \frac{\partial \mathbf X}{\partial q^i}, \quad
G^{ij} = \frac{\partial q^i}{\partial \mathbf X} \cdot \frac{\partial q^j}{\partial \mathbf X}.
\end{equation}

We now wish to define the conformation tensor $\mathbf A$ whose eigenvalues give the stretches of the polymer. This is in direct analogy to Finger tensor $\mathbf B$, the only difference being that the stretches need to be measured with respect to the instantaneous reference state $\mathbf X(q^i,t)$, rather than  $\mathbf x(q^i,t=0)$. Hence, we can define

\begin{equation}
\mathbf A = \left( \frac{\partial q^i}{\partial \mathbf X} \cdot \frac{\partial q^j}{\partial \mathbf X} \right) \mathbf e_i \otimes \mathbf e_j,
\end{equation}
where now the components $A^{ij}$ are time-dependent due to the
relaxation of $\mathbf X(q^i,t)$. If we wish to express this relaxation
directly in terms of the  conformation tensor $\mathbf A$, we arrive
at
\begin{equation}
\frac{dA^{ij}}{dt} = \frac{1}{\lambda} f^{ij}(\mathbf A),
\end{equation}
where the right hand side is {\it independent} of the flow, since any
time dependence only enters through the reference state $\mathbf X(q^i,t)$.
By contrast, the covariant components $A_{ij} = g_{ik}g_{jm} A^{km}$
exhibit a time-dependence due to the flow $dg_{ij}/dt$ and due to the
relaxation of $\mathbf X(q^i,t)$. Hence, to quantify the relaxation
in terms of the conformation tensor $\mathbf A$, one automatically
singles out the upper convected derivatives as the appropriate time derivative.

In analogy with elasticity theory, we introduce an elastic energy $W(I_1,I_2)$ associated to the first and second invariants of $\mathbf A$. Once again we employ the virtual work principle, but now including dissipation $\epsilon_p$:

\begin{equation}\label{eq:virtualbis}
\frac{1}{2}\sigma^{ij} \dot{\gamma}_{ij} = \frac{dW}{dt} +\epsilon_p.
\end{equation}
The dissipation is necessary since the elastic energy exhibit an extra time-dependence associated to relaxation,

\begin{equation}
\frac{dW}{dt} =\frac{\partial W}{\partial g_{ij}}  \dot{\gamma}_{ij}  + \frac{\partial W}{\partial A^{ij}} \frac{dA^{ij}}{dt}, \end{equation}
Again, the terms proportional to $\dot{\gamma}_{ij}$ provide the stress, so that stress and dissipation can be separated as

\begin{equation}
\sigma^{ij}=2\frac{\partial W}{\partial g_{ij}}, \quad \quad \epsilon_p = - \frac{\partial W}{\partial A^{ij}} \frac{dA^{ij}}{dt}.
\end{equation}


\begin{thebibliography}{41}
\expandafter\ifx\csname natexlab\endcsname\relax\def\natexlab#1{#1}\fi
\expandafter\ifx\csname bibnamefont\endcsname\relax
  \def\bibnamefont#1{#1}\fi
\expandafter\ifx\csname bibfnamefont\endcsname\relax
  \def\bibfnamefont#1{#1}\fi
\expandafter\ifx\csname citenamefont\endcsname\relax
  \def\citenamefont#1{#1}\fi
\expandafter\ifx\csname url\endcsname\relax
  \def\url#1{\texttt{#1}}\fi
\expandafter\ifx\csname urlprefix\endcsname\relax\def\urlprefix{URL }\fi
\providecommand{\bibinfo}[2]{#2}
\providecommand{\eprint}[2][]{\url{#2}}

\bibitem[{\citenamefont{Bird et~al.}(1987)\citenamefont{Bird, Armstrong, and
  Hassager}}]{BAH87}
\bibinfo{author}{\bibfnamefont{R.~B.} \bibnamefont{Bird}},
  \bibinfo{author}{\bibfnamefont{R.~C.} \bibnamefont{Armstrong}},
  \bibnamefont{and} \bibinfo{author}{\bibfnamefont{O.}~\bibnamefont{Hassager}},
  \emph{\bibinfo{title}{Dynamics of Polymeric Liquids Volume I: Fluid
  Mechanics; Volume II: Kinetic Theory}} (\bibinfo{publisher}{Wiley: New York},
  \bibinfo{year}{1987}).

\bibitem[{\citenamefont{Larson}(1999)}]{Larson99}
\bibinfo{author}{\bibfnamefont{R.~G.} \bibnamefont{Larson}},
  \emph{\bibinfo{title}{The structure and rhology of complex fluids}}
  (\bibinfo{publisher}{Oxford University Press}, \bibinfo{address}{Oxford, UK},
  \bibinfo{year}{1999}).

\bibitem[{\citenamefont{Morozov and Spagnolie}(2015)}]{MS15}
\bibinfo{author}{\bibfnamefont{A.}~\bibnamefont{Morozov}} \bibnamefont{and}
  \bibinfo{author}{\bibfnamefont{S.~E.} \bibnamefont{Spagnolie}},
  \emph{\bibinfo{title}{Introduction to complex fluids}}
  (\bibinfo{publisher}{Springer}, \bibinfo{year}{2015}).

\bibitem[{\citenamefont{Clasen et~al.}(2006)\citenamefont{Clasen, Eggers,
  Fontelos, Li, and McKinley}}]{CEFLM06}
\bibinfo{author}{\bibfnamefont{C.}~\bibnamefont{Clasen}},
  \bibinfo{author}{\bibfnamefont{J.}~\bibnamefont{Eggers}},
  \bibinfo{author}{\bibfnamefont{M.~A.} \bibnamefont{Fontelos}},
  \bibinfo{author}{\bibfnamefont{J.}~\bibnamefont{Li}}, \bibnamefont{and}
  \bibinfo{author}{\bibfnamefont{G.~H.} \bibnamefont{McKinley}},
  \bibinfo{journal}{J. Fluid Mech.} \textbf{\bibinfo{volume}{556}},
  \bibinfo{pages}{283} (\bibinfo{year}{2006}).

\bibitem[{\citenamefont{Morozov and {van Saarloos}}(2007)}]{MvS07}
\bibinfo{author}{\bibfnamefont{A.~N.} \bibnamefont{Morozov}} \bibnamefont{and}
  \bibinfo{author}{\bibfnamefont{W.}~\bibnamefont{{van Saarloos}}},
  \bibinfo{journal}{Phys. Rep.} \textbf{\bibinfo{volume}{447}},
  \bibinfo{pages}{112} (\bibinfo{year}{2007}).

\bibitem[{\citenamefont{Martin et~al.}(1972)\citenamefont{Martin, Parodi, and
  Pershan}}]{MPP72}
\bibinfo{author}{\bibfnamefont{P.~C.} \bibnamefont{Martin}},
  \bibinfo{author}{\bibfnamefont{O.}~\bibnamefont{Parodi}}, \bibnamefont{and}
  \bibinfo{author}{\bibfnamefont{P.~S.} \bibnamefont{Pershan}},
  \bibinfo{journal}{Phys. Rev. A} \textbf{\bibinfo{volume}{8}},
  \bibinfo{pages}{2401} (\bibinfo{year}{1972}).

\bibitem[{\citenamefont{Leonov}(1976)}]{Leonov1976}
\bibinfo{author}{\bibfnamefont{A.~I.} \bibnamefont{Leonov}},
  \bibinfo{journal}{Rheologica Acta} \textbf{\bibinfo{volume}{15}},
  \bibinfo{pages}{85} (\bibinfo{year}{1976}).

\bibitem[{\citenamefont{Beris and Edwards}(1994)}]{BE_book}
\bibinfo{author}{\bibfnamefont{A.~N.} \bibnamefont{Beris}} \bibnamefont{and}
  \bibinfo{author}{\bibfnamefont{B.~J.} \bibnamefont{Edwards}},
  \emph{\bibinfo{title}{Thermodynamics of flowing systems}}
  (\bibinfo{publisher}{Oxford University Press}, \bibinfo{year}{1994}).

\bibitem[{\citenamefont{{\"O}ttinger}(2005)}]{Oe_book}
\bibinfo{author}{\bibfnamefont{H.~C.} \bibnamefont{{\"O}ttinger}},
  \emph{\bibinfo{title}{Beyond equilibrium thermodynamics}}
  (\bibinfo{publisher}{Wiley-Interscience}, \bibinfo{year}{2005}).

\bibitem[{\citenamefont{Tanner}(2000)}]{BookTanner}
\bibinfo{author}{\bibfnamefont{R.~I.} \bibnamefont{Tanner}},
  \emph{\bibinfo{title}{Engineering Rheology}} (\bibinfo{publisher}{Oxford
  University Press}, \bibinfo{year}{2000}).

\bibitem[{\citenamefont{Temmen et~al.}(2000)\citenamefont{Temmen, Pleiner, Liu,
  and Brand}}]{TPLB00}
\bibinfo{author}{\bibfnamefont{H.}~\bibnamefont{Temmen}},
  \bibinfo{author}{\bibfnamefont{H.}~\bibnamefont{Pleiner}},
  \bibinfo{author}{\bibfnamefont{M.}~\bibnamefont{Liu}}, \bibnamefont{and}
  \bibinfo{author}{\bibfnamefont{H.~R.} \bibnamefont{Brand}},
  \bibinfo{journal}{Phys. Rev. Lett.} \textbf{\bibinfo{volume}{84}},
  \bibinfo{pages}{3228} (\bibinfo{year}{2000}).

\bibitem[{\citenamefont{Beris et~al.}(2001)\citenamefont{Beris, Graham, Karlin,
  and {\"O}ttinger}}]{BGKO01}
\bibinfo{author}{\bibfnamefont{A.~N.} \bibnamefont{Beris}},
  \bibinfo{author}{\bibfnamefont{M.~D.} \bibnamefont{Graham}},
  \bibinfo{author}{\bibfnamefont{I.}~\bibnamefont{Karlin}}, \bibnamefont{and}
  \bibinfo{author}{\bibfnamefont{H.~C.} \bibnamefont{{\"O}ttinger}},
  \bibinfo{journal}{Phys. Rev. Lett.} \textbf{\bibinfo{volume}{86}},
  \bibinfo{pages}{744} (\bibinfo{year}{2001}).

\bibitem[{\citenamefont{M{\"u}ller et~al.}(2016)\citenamefont{M{\"u}ller, Liu,
  Pleiner, and Brand}}]{MLPB16b}
\bibinfo{author}{\bibfnamefont{O.}~\bibnamefont{M{\"u}ller}},
  \bibinfo{author}{\bibfnamefont{M.}~\bibnamefont{Liu}},
  \bibinfo{author}{\bibfnamefont{H.}~\bibnamefont{Pleiner}}, \bibnamefont{and}
  \bibinfo{author}{\bibfnamefont{H.~R.} \bibnamefont{Brand}},
  \bibinfo{journal}{Phys. Rev. E} \textbf{\bibinfo{volume}{93}},
  \bibinfo{pages}{023114} (\bibinfo{year}{2016}).

\bibitem[{\citenamefont{Mora et~al.}(2010)\citenamefont{Mora, Phou, Frometal,
  Pismen, and Pomeau}}]{MPFPP10}
\bibinfo{author}{\bibfnamefont{S.}~\bibnamefont{Mora}},
  \bibinfo{author}{\bibfnamefont{T.}~\bibnamefont{Phou}},
  \bibinfo{author}{\bibfnamefont{J.-M.} \bibnamefont{Frometal}},
  \bibinfo{author}{\bibfnamefont{L.~M.} \bibnamefont{Pismen}},
  \bibnamefont{and} \bibinfo{author}{\bibfnamefont{Y.}~\bibnamefont{Pomeau}},
  \bibinfo{journal}{Phys. Rev. Lett.} \textbf{\bibinfo{volume}{105}},
  \bibinfo{pages}{214301} (\bibinfo{year}{2010}).

\bibitem[{\citenamefont{Andreotti et~al.}(2016)\citenamefont{Andreotti,
  B{\"a}umchen, Boulogne, Daniels, Dufresne, Perrin, Salez, Snoeijer, and
  Style}}]{AndreottiSoftMatt2016}
\bibinfo{author}{\bibfnamefont{B.}~\bibnamefont{Andreotti}},
  \bibinfo{author}{\bibfnamefont{O.}~\bibnamefont{B{\"a}umchen}},
  \bibinfo{author}{\bibfnamefont{F.}~\bibnamefont{Boulogne}},
  \bibinfo{author}{\bibfnamefont{K.~E.} \bibnamefont{Daniels}},
  \bibinfo{author}{\bibfnamefont{E.~R.} \bibnamefont{Dufresne}},
  \bibinfo{author}{\bibfnamefont{H.}~\bibnamefont{Perrin}},
  \bibinfo{author}{\bibfnamefont{T.}~\bibnamefont{Salez}},
  \bibinfo{author}{\bibfnamefont{J.~H.} \bibnamefont{Snoeijer}},
  \bibnamefont{and} \bibinfo{author}{\bibfnamefont{R.~W.} \bibnamefont{Style}},
  \bibinfo{journal}{Soft Matter} \textbf{\bibinfo{volume}{12}},
  \bibinfo{pages}{2993} (\bibinfo{year}{2016}).

\bibitem[{\citenamefont{Style et~al.}(2017)\citenamefont{Style, Jagota, Hui,
  and Dufresne}}]{StyleDufresneReview}
\bibinfo{author}{\bibfnamefont{R.~W.} \bibnamefont{Style}},
  \bibinfo{author}{\bibfnamefont{A.}~\bibnamefont{Jagota}},
  \bibinfo{author}{\bibfnamefont{C.}~\bibnamefont{Hui}}, \bibnamefont{and}
  \bibinfo{author}{\bibfnamefont{E.~R.} \bibnamefont{Dufresne}},
  \bibinfo{journal}{Annual Review of Condensed Matter Physics}
  \textbf{\bibinfo{volume}{8}}, \bibinfo{pages}{99} (\bibinfo{year}{2017}).

\bibitem[{\citenamefont{Entov and Yarin}(1984)}]{EYa84}
\bibinfo{author}{\bibfnamefont{V.~M.} \bibnamefont{Entov}} \bibnamefont{and}
  \bibinfo{author}{\bibfnamefont{A.~L.} \bibnamefont{Yarin}},
  \bibinfo{journal}{Fluid Dyn.} \textbf{\bibinfo{volume}{19}},
  \bibinfo{pages}{21} (\bibinfo{year}{1984}).

\bibitem[{\citenamefont{Eggers and Fontelos}(2015)}]{EF_book}
\bibinfo{author}{\bibfnamefont{J.}~\bibnamefont{Eggers}} \bibnamefont{and}
  \bibinfo{author}{\bibfnamefont{M.~A.} \bibnamefont{Fontelos}},
  \emph{\bibinfo{title}{Singularities: Formation, Structure, and Propagation}}
  (\bibinfo{publisher}{Cambridge University Press, Cambridge},
  \bibinfo{year}{2015}).

\bibitem[{\citenamefont{Xuan and Biggins}(2017)}]{XB17}
\bibinfo{author}{\bibfnamefont{C.}~\bibnamefont{Xuan}} \bibnamefont{and}
  \bibinfo{author}{\bibfnamefont{J.}~\bibnamefont{Biggins}},
  \bibinfo{journal}{Phys. Rev. E} \textbf{\bibinfo{volume}{95}},
  \bibinfo{pages}{053106} (\bibinfo{year}{2017}).

\bibitem[{\citenamefont{Mwasame et~al.}(2017)\citenamefont{Mwasame, Wagner, and
  Beris}}]{MWB17}
\bibinfo{author}{\bibfnamefont{P.~M.} \bibnamefont{Mwasame}},
  \bibinfo{author}{\bibfnamefont{N.~J.} \bibnamefont{Wagner}},
  \bibnamefont{and} \bibinfo{author}{\bibfnamefont{A.~N.} \bibnamefont{Beris}},
  \bibinfo{journal}{J. Fluid Mech.} \textbf{\bibinfo{volume}{831}},
  \bibinfo{pages}{433} (\bibinfo{year}{2017}).

\bibitem[{\citenamefont{Mwasame et~al.}(2018)\citenamefont{Mwasame, Wagner, and
  Beris}}]{MWB18}
\bibinfo{author}{\bibfnamefont{P.~M.} \bibnamefont{Mwasame}},
  \bibinfo{author}{\bibfnamefont{N.~J.} \bibnamefont{Wagner}},
  \bibnamefont{and} \bibinfo{author}{\bibfnamefont{A.~N.} \bibnamefont{Beris}},
  \bibinfo{journal}{Phys. Fluids} \textbf{\bibinfo{volume}{30}},
  \bibinfo{pages}{030704} (\bibinfo{year}{2018}).

\bibitem[{\citenamefont{Oldroyd}(1950)}]{O50}
\bibinfo{author}{\bibfnamefont{J.~G.} \bibnamefont{Oldroyd}},
  \bibinfo{journal}{Proc. Roy. Soc. A} \textbf{\bibinfo{volume}{200}},
  \bibinfo{pages}{523} (\bibinfo{year}{1950}).

\bibitem[{\citenamefont{Beris and Mavrantzas}(1994)}]{BM94}
\bibinfo{author}{\bibfnamefont{A.~N.} \bibnamefont{Beris}} \bibnamefont{and}
  \bibinfo{author}{\bibfnamefont{V.~G.} \bibnamefont{Mavrantzas}},
  \bibinfo{journal}{J. Rheol.} \textbf{\bibinfo{volume}{38}},
  \bibinfo{pages}{1235} (\bibinfo{year}{1994}).

\bibitem[{\citenamefont{Landau and Lifshitz}(1984)}]{LL7}
\bibinfo{author}{\bibfnamefont{L.~D.} \bibnamefont{Landau}} \bibnamefont{and}
  \bibinfo{author}{\bibfnamefont{E.~M.} \bibnamefont{Lifshitz}},
  \emph{\bibinfo{title}{Elasticity}} (\bibinfo{publisher}{Pergamon: Oxford},
  \bibinfo{year}{1984}).

\bibitem[{Wik()}]{Wiki_notes_FST}
\urlprefix\url{https://en.wikipedia.org/wiki/Finite_strain_theory}.

\bibitem[{\citenamefont{Truesdell and Noll}(2004)}]{TN_book}
\bibinfo{author}{\bibfnamefont{C.}~\bibnamefont{Truesdell}} \bibnamefont{and}
  \bibinfo{author}{\bibfnamefont{W.}~\bibnamefont{Noll}},
  \emph{\bibinfo{title}{The Non-Linear Field Theories of Mechanics}},
  \bibinfo{number}{v. 3} (\bibinfo{publisher}{Springer}, \bibinfo{year}{2004}).

\bibitem[{\citenamefont{Kamrin et~al.}(2012)\citenamefont{Kamrin, Rycroft, and
  Nave}}]{Kamrin2012}
\bibinfo{author}{\bibfnamefont{K.}~\bibnamefont{Kamrin}},
  \bibinfo{author}{\bibfnamefont{C.~H.} \bibnamefont{Rycroft}},
  \bibnamefont{and} \bibinfo{author}{\bibfnamefont{J.~C.} \bibnamefont{Nave}},
  \bibinfo{journal}{Journal of the Mechanics and Physics of Solids}
  \textbf{\bibinfo{volume}{60}}, \bibinfo{pages}{1952} (\bibinfo{year}{2012}).

\bibitem[{\citenamefont{Magnus and Neudecker}(1999)}]{Magnus_Neudecker}
\bibinfo{author}{\bibfnamefont{J.~R.} \bibnamefont{Magnus}} \bibnamefont{and}
  \bibinfo{author}{\bibfnamefont{H.}~\bibnamefont{Neudecker}},
  \emph{\bibinfo{title}{Matrix Differential Calculus with Applications in
  Statistics and Econometrics, 2nd Edition}} (\bibinfo{publisher}{Wiley},
  \bibinfo{year}{1999}).

\bibitem[{\citenamefont{{J}ohnson {J}r. and Segalman}(1977)}]{JS77}
\bibinfo{author}{\bibfnamefont{M.~W.} \bibnamefont{{J}ohnson {J}r.}}
  \bibnamefont{and} \bibinfo{author}{\bibfnamefont{D.}~\bibnamefont{Segalman}},
  \bibinfo{journal}{J. Non-Newtonian Fluid Mech.} \textbf{\bibinfo{volume}{2}},
  \bibinfo{pages}{255} (\bibinfo{year}{1977}).

\bibitem[{\citenamefont{Hohenegger and Shelley}(2011)}]{HS_les_houches}
\bibinfo{author}{\bibfnamefont{C.}~\bibnamefont{Hohenegger}} \bibnamefont{and}
  \bibinfo{author}{\bibfnamefont{M.}~\bibnamefont{Shelley}},
  \emph{\bibinfo{title}{Dynamics of complex biofluids}}
  (\bibinfo{publisher}{Oxford University Press}, \bibinfo{year}{2011}),
  vol.~\bibinfo{volume}{92}.

\bibitem[{\citenamefont{Chilcott and Rallison}(1988)}]{CR88}
\bibinfo{author}{\bibfnamefont{M.~D.} \bibnamefont{Chilcott}} \bibnamefont{and}
  \bibinfo{author}{\bibfnamefont{J.~M.} \bibnamefont{Rallison}},
  \bibinfo{journal}{J. Non-Newtonian Fluid Mech.}
  \textbf{\bibinfo{volume}{29}}, \bibinfo{pages}{381} (\bibinfo{year}{1988}).

\bibitem[{\citenamefont{Mihai and Goriely}(2017)}]{Mihai17}
\bibinfo{author}{\bibfnamefont{L.~A.} \bibnamefont{Mihai}} \bibnamefont{and}
  \bibinfo{author}{\bibfnamefont{A.}~\bibnamefont{Goriely}},
  \bibinfo{journal}{Proceedings of the Royal Society A: Mathematical, Physical
  and Engineering Sciences} \textbf{\bibinfo{volume}{473}},
  \bibinfo{pages}{20170607} (\bibinfo{year}{2017}).

\bibitem[{\citenamefont{Turkoz et~al.}(2018)\citenamefont{Turkoz,
  {Lopez-Herrera}, Eggers, Arnold, and Deike}}]{TLEAD18}
\bibinfo{author}{\bibfnamefont{E.}~\bibnamefont{Turkoz}},
  \bibinfo{author}{\bibfnamefont{J.~M.} \bibnamefont{{Lopez-Herrera}}},
  \bibinfo{author}{\bibfnamefont{J.}~\bibnamefont{Eggers}},
  \bibinfo{author}{\bibfnamefont{C.~B.} \bibnamefont{Arnold}},
  \bibnamefont{and} \bibinfo{author}{\bibfnamefont{L.}~\bibnamefont{Deike}},
  \bibinfo{journal}{J. Fluid Mech.} \textbf{\bibinfo{volume}{851}},
  \bibinfo{pages}{R2} (\bibinfo{year}{2018}).

\bibitem[{\citenamefont{Bazilevskii et~al.}(1997)\citenamefont{Bazilevskii,
  Entov, Lerner, and Rozhkov}}]{BELR97}
\bibinfo{author}{\bibfnamefont{A.~V.} \bibnamefont{Bazilevskii}},
  \bibinfo{author}{\bibfnamefont{V.~M.} \bibnamefont{Entov}},
  \bibinfo{author}{\bibfnamefont{M.~M.} \bibnamefont{Lerner}},
  \bibnamefont{and} \bibinfo{author}{\bibfnamefont{A.~N.}
  \bibnamefont{Rozhkov}}, \bibinfo{journal}{Polym. Sci. Ser. A}
  \textbf{\bibinfo{volume}{39}}, \bibinfo{pages}{316} (\bibinfo{year}{1997}).

\bibitem[{\citenamefont{Herrada and Montanero}(2016)}]{Herrada2016}
\bibinfo{author}{\bibfnamefont{M.}~\bibnamefont{Herrada}} \bibnamefont{and}
  \bibinfo{author}{\bibfnamefont{J.}~\bibnamefont{Montanero}},
  \bibinfo{journal}{J. Comp. Phys.} \textbf{\bibinfo{volume}{306}}
  (\bibinfo{year}{2016}).

\bibitem[{\citenamefont{Negahban}(2012)}]{Negahban12}
\bibinfo{author}{\bibfnamefont{M.}~\bibnamefont{Negahban}},
  \emph{\bibinfo{title}{The mechanical and thermodynamical theory of
  plasticity}} (\bibinfo{publisher}{CRC Press}, \bibinfo{year}{2012}).

\bibitem[{\citenamefont{Eggers et~al.}(2019)\citenamefont{Eggers, Snoeijer, and
  Herrada}}]{EHS19}
\bibinfo{author}{\bibfnamefont{J.}~\bibnamefont{Eggers}},
  \bibinfo{author}{\bibfnamefont{J.~H.} \bibnamefont{Snoeijer}},
  \bibnamefont{and} \bibinfo{author}{\bibfnamefont{M.~A.}
  \bibnamefont{Herrada}}, \bibinfo{journal}{Preprint}  (\bibinfo{year}{2019}).

\bibitem[{\citenamefont{Eggers and Villermaux}(2008)}]{EV08}
\bibinfo{author}{\bibfnamefont{J.}~\bibnamefont{Eggers}} \bibnamefont{and}
  \bibinfo{author}{\bibfnamefont{E.}~\bibnamefont{Villermaux}},
  \bibinfo{journal}{Rep. Progr. Phys.} \textbf{\bibinfo{volume}{71}},
  \bibinfo{pages}{036601} (\bibinfo{year}{2008}).

\bibitem[{\citenamefont{Green and Zerna}(2002)}]{green2002}
\bibinfo{author}{\bibfnamefont{A.}~\bibnamefont{Green}} \bibnamefont{and}
  \bibinfo{author}{\bibfnamefont{W.}~\bibnamefont{Zerna}},
  \emph{\bibinfo{title}{Theoretical Elasticity}}, Phoenix Edition Series
  (\bibinfo{publisher}{Dover Publications}, \bibinfo{year}{2002}).

\bibitem[{\citenamefont{Hanna}(2019)}]{Hanna19}
\bibinfo{author}{\bibfnamefont{J.~A.} \bibnamefont{Hanna}},
  \bibinfo{journal}{arXiv:1807.06426v3 [cond-mat.soft]}
  (\bibinfo{year}{2019}).

\bibitem[{\citenamefont{Aris}(1990)}]{aris1990}
\bibinfo{author}{\bibfnamefont{R.}~\bibnamefont{Aris}},
  \emph{\bibinfo{title}{Vectors, Tensors and the Basic Equations of Fluid
  Mechanics}}, Dover Books on Mathematics (\bibinfo{publisher}{Dover
  Publications}, \bibinfo{year}{1990}).

\end{thebibliography}

\end{document}